\documentclass[prd,showpacs,preprint,eqsecnum,amsmath,amssymb,nofootinbib]{revtex4}
\usepackage{graphicx}
\usepackage{amssymb,amsmath}
\newcommand{\be}{\begin{equation}}
\newcommand{\ee}{\end{equation}}
\newcommand{\bea}{\begin{eqnarray}}
\newcommand{\eea}{\end{eqnarray}}
\newcommand{\bes}{\begin{subequations}}
\newcommand{\ees}{\end{subequations}}

\begin{document}
\title{Hawking radiation correlations in Bose Einstein condensates \\ using quantum field theory in curved space}

\author{Paul~R.~Anderson}
\email{anderson@wfu.edu}
\affiliation{Department of Physics, Wake Forest University, Winston-Salem, North Carolina 27109, USA}
\author{Roberto~Balbinot}
\email{Roberto.Balbinot@bo.infn.it}
\affiliation{Dipartimento di Fisica dell'Universit\'a di Bologna and INFN sezione di Bologna, Via Irnerio 46, 40126 Bologna, Italy}
\author{Alessandro~Fabbri}
\email{afabbri@ific.uv.es}
\affiliation{Museo Storico della Fisica e Centro Studi e Ricerche 'Enrico Fermi', \\
Piazza del Viminale 1, 00184 Roma, Italy; \\
Dipartimento di Fisica dell'Universit\'a di Bologna,\\
Via Irnerio 46, 40126 Bologna, Italy;\\
Departamento de F\'isica Te\'orica and IFIC, Universidad de Valencia-CSIC,\\
C. Dr. Moliner 50, 46100 Burjassot, Spain}
\author{Renaud~Parentani}
\email{renaud.parentani@th.u-psud.fr}
\affiliation{Laboratoire de Physique Th\'eorique, CNRS UMR 8627, B\^at. 210, Universit\'e Paris-Sud 11, 91405 Orsay Cedex, France}

\begin{abstract}
The density density correlation function is computed for the Bogoliubov pseudoparticles created in a Bose-Einstein condensate undergoing a black hole flow. On the basis of the gravitational analogy, the method used relies only on quantum field theory in curved spacetime techniques.  A comparison with the results obtained by ab initio full condensed matter calculations is given, confirming the validity of the approximation used provided the profile of the flow varies smoothly on scales compared to the condensate healing length.
\end{abstract}
\maketitle
\section{introduction}
It is by now well known that the behavior of elementary excitations in various condensed matter systems can be described, under some approximation, by an effective theory of massless scalars (phonons) propagating on a fictitious curved spacetime according to a covariant Klein-Gordon equation \cite{Unruh,lrr}. This curved spacetime, which
has nothing to do with the real physical spacetime on which the underlying condensed matter theory is defined, is determined by the background features of the condensed matter system. In particular, curvature is associated with inhomogeneities of the system. The approximation, which is at the core of this condensed-matter analogy, is typically a hydrodynamical approximation whose validity requires one to consider the system on scales much larger than the typical length associated with the microscopic (atomic) structure of the system. \par
It is of particular interest to consider configurations for which the associated curved spacetime metric describes what in gravity would be called a black hole, BH.  This happens, for example, in a stationary flow when the speed of sound varies with position in such a way that there is both a subsonic and a supersonic region with the direction of flow from the subsonic towards the supersonic region.  The surface separating the subsonic from the supersonic regime acts as a sort of ``acoustic horizon'' since sound waves in the supersonic region (the ''acoustic'' BH) are trapped inside this horizon being unable to propagate upstream towards the subsonic part, mimicking the trapping of light (and everything else) inside the horizon of a gravitational BH. \par
There is therefore the concrete possibility of investigating in these condensed matter configurations the analogue of one of the more exotic processes foreseen in theoretical physics, namely BH quantum evaporation as predicted by Hawking in 1974 ~\cite{hawking}. This process, despite what one would naively think, is not peculiar to gravity. It is completely kinematical as it relies only on the presence of a horizon in the underlying metric on which the quantum fields propagate. As such it is expected to occur in condensed matter analogue  systems when the associated curved spacetime has a BH form~\cite{Unruh,lrr}. \par
Hawking evaporation is a general pair creation process in which quantum vacuum fluctuations are converted into real on shell quanta. One quantum (the positive energy one) is emitted outside the horizon and propagates to infinity where it is characterized by a thermal spectrum at the Hawking temperature, the other member of the pair (the negative Killing energy quantum), called its ``partner'' \cite{mapar}, propagates inside the horizon and remains trapped there. It is this compensation of positive and negative energy which allows the process to occur in stationary, even static configurations. \par
For a gravitational BH the Hawking temperature associated with this emission process is extremely low $T_H\sim 10^{-7}(M_{Sun}/M) \, K$, where $M$ is the BH mass.  This makes Hawking evaporation astrophysically irrelevant and impossible to detect for stellar mass or larger black holes since the signal is completely overwhelmed by other sources including the $2.7\ K$ cosmic microwave background radiation.  In principle Hawking radiation from primordial BHs~\cite{pm1,pm2} with small enough masses could be detected~\cite{det1,det2}, but no BHs of this type have been discovered.\par
 Thus analogue systems  appear today as the only realistic hope of finding experimental verification of Hawking's prediction. But even in this context the experimental situation is difficult due to competing effects such as thermal fluctuations and quantum noise.  These effects can mask the signal even in the most favorable situations where the associated Hawking temperature is as high as $1/10$ of the background temperature.\par
A way to bypass this critical problem was proposed in~\cite{paper1} where it was shown that the Hawking process encodes characteristic correlations between the quanta and their partners that can be measured in analogue systems, allowing for clear identification of the Hawking process.  This kind of in-out correlation is not of physical interest for a gravitational BH since external observers have no access to the interior BH region (where the partners live).  For acoustic BHs the region inside the horizon is simply the supersonic part of a flow and hence is physically accessible for measurements.  The relationships between the equal-time correlations studied in analogue BHs~\cite{paper1,paper2} and
the space-time correlation pattern found in relativistic theories are discussed in~\cite{From2010}. \par
One of the most highly studied systems which can be used to create an analogue black hole is a Bose-Einstein condensate~\cite{pistri}, BEC, which is an ensemble of sufficiently cold atoms, the vast majority of which are in the ground state of the system.  The small fraction of the atoms which are in excited states can be described as (quantum) excitations above a classical ($c$ number) condensate field.

In Ref.~\cite{paper1}, using the condensed matter - gravity analogy, the correlations associated with Hawking radiation in a BEC elongated in one spatial direction undergoing a BH-like flow were computed using the standard tools of quantum field theory, QFT, in curved space.  Two approximations were made to simplify the calculations.  The first was to reduce the mode equation from a 4-D form to a 2-D form.  In the dimensional reduction process there is a potential that appears in the 2-D mode equation which causes backscattering of the modes to occur.  The second approximation was to neglect this potential.  Without it the modes propagate freely as 2-D plane waves. \par
It was found that if a sonic horizon is present then there is a negative correlation peak in the density density correlation function when one point is inside and one point is outside the horizon.  No such peak occurs if there is no horizon.  In~\cite{paper2} ab initio full condensed matter calculations of the density-density correlation function were made.  The calculations were based on Monte Carlo simulations within the full microscopic quantum description of the Bose-Einstein condensate.
Comparisons were made with the analytic calculation in~\cite{paper1} and it was found that the two calculations are in approximate agreement for both the size of the peak and the width of the peak so long as the flow varies smoothly on scales compared to the condensate healing length. \par
The analytic calculation in~\cite{paper1} was carried out only for the case in which one point is inside and one point is outside the horizon.  However, it can be generalized in a straight-forward way to consider all possible pairings of the two points.  When this is done the only other structure found is the usual peak that occurs when the two points come together.
 The calculation of the density density correlation function in~\cite{paper2} showed a richer structure.  In particular a negative correlation peak was found in the case that both points are inside the horizon.  This is a second effect which depends on the existence of a horizon.  A third, positive correlation peak was also found~\cite{paper2, rpc, paper3} when one point is inside and one point is outside the horizon.  It was found that this peak occurs even if the horizon is not present. \par
Because the calculation in~\cite{paper1} neglected the potential in the 2-D mode equation it is natural to ask whether this potential has any important effects.  In this paper we answer that question by numerically computing the density density correlation function when the potential is included in the mode equation.  We find that the resulting backscattering of the modes leads to a more complex structure for the correlation pattern which fits very well with the one obtained in~\cite{paper2}.  \par
In Sec.\ II the relevant properties and equations for a BEC are reviewed.  In Sec.\ III the specific model we consider is discussed.  In Sec.\ IV the Unruh state which we use for our calculations is discussed and formulas for the relevant Bogolubov coefficients are derived.  In Sec.\ V the derivation of the specific expressions for the density density correlation function that are used in our numerical calculations are given.  Sec.\ VI contains a derivation of the specific mode equations which we solve. In Sec.\ VII the numerical computation of the density density correlation function is discussed.  The results of the numerical calculations are given in Sec.\ VIII and our conclusions are given in Sec.\ IX.

\section{Bose-Einstein condensates and the associated analogue metric}

Here we shall briefly review the basic equations describing a BEC and how these lead, under the hydrodynamical approximation, to the context of an acoustic metric effectively governing the propagation of the fluctuations.\par
In the Bogoliubov theory the basic bosonic-field operator $\hat\Psi$ describing the atoms is splitted in a classical mean field $\Psi_0$ describing the macroscopic occupied low energy state of the system (the condensate) and a part describing the quantum fluctuation above this classical state
\be \hat\Psi =\Psi_0 (1+\hat\phi)\ . \ee
The evolution of the condensate is governed by the Gross-Pitaevskii equation
\begin{equation}\label{gp}
i\hbar \partial_T \Psi_0 = \left(-\frac{\hbar^2}{2m}\vec \nabla^2 + V_{ext} + g |\Psi_0|^2 \right)\,\Psi_0\ ,
\end{equation}
where $m$ is the mass of the atoms, $g$ the coupling and $V_{ext}$ the external potential confining the atoms. The field operator $\hat\phi$ describing the noncondensed part satisfies the Bogoliubov-de Gennes, BdG, equation
\begin{equation}\label{bdg}
i\hbar \partial_T \hat \phi= - \left( \frac{\hbar^2 \vec \nabla^2}{2m} + \frac{\hbar^2}{m}\frac{\vec \nabla \Psi_0 }{\Psi_0} \vec \nabla\right)\hat\phi +ng (\hat\phi + \hat\phi^{\dagger}),
\end{equation}
where $n\equiv |\Psi_0|^2$ is the condensate density.\par
The BdG equation  (\ref{bdg})  can be manipulated to obtain a curved space wave equation as follows. One rewrites the condensate wave function $\Psi_0$ and the basic field operator $\hat\Psi$ in the so called density-phase representation, namely
\bea && \Psi_0 = \sqrt{n}\- e^{i\theta}\ , \nonumber \\ && \hat\Psi = \sqrt{n+\hat n_1}\  e^{i(\theta + \hat \theta_1)}\simeq \Psi_0 (1+ \frac{\hat n_1}{2n} +i\hat\theta_1 )\ .\eea
In this representation the BdG equation becomes a pair of equations of motion for the density fluctuation $\hat n_1$ and the phase fluctuation $\hat \theta_1$
\begin{eqnarray}\label{bodg1}
&&\hbar\partial_T \hat\theta_1=-\hbar \vec v_0\vec \nabla \hat \theta_1 - \frac{mc^2}{n}\hat n_1
 + \frac{mc^2}{4n}\xi^2 \vec \nabla [ n\vec \nabla(\frac{\hat n_1}{n})]=0
 \ ,\\ \label{bodg2}
&&\partial_T\hat n_1=-\vec \nabla (\vec v_0 \hat n_1+ \frac{\hbar n}{m}\vec \nabla\theta_1 )\  ,
\end{eqnarray}
where $\vec v_0=\hbar\,\nabla\vec \theta / m$ is the condensate flow velocity and $c=\sqrt{ng/m}$ the local speed of
sound which will play a critical role in our construction. $\xi\equiv \hbar/mc$  is the so called healing length,
it is the fundamental scale in describing the microscopic quantum structure of the BEC.\par
If we limit ourselves to scales much bigger than $\xi$, i.e. within the so called hydrodynamical approximation, the last term in Eq.~(\ref{bodg1}) can be neglected namely \begin{equation}\label{bhu}
\hat n_1 \simeq -\frac{\hbar n}{mc^2}\left[ \vec v_0\vec \nabla\hat\theta_1 +\partial_T\hat\theta_1\right] \;.
\end{equation}
Substituting this equation into Eq.~\eqref{bodg2} then decouples the system of equations (\ref{bodg1}, \ref{bodg2}) yielding the equation
\begin{equation}\label{eqfase}
-(\partial_T+\vec \nabla\vec v_0)\frac{n}{mc^2}(\partial_T+\vec v_0\vec \nabla)\theta_1 + \vec \nabla\frac{n}{m}\vec \nabla\theta_1=0 \;,
\end{equation}
which can formally be rewritten as
\be
\hat \Box \hat\theta_1 =0\ ,
\ee
where
\begin{equation}\label{box}
\Box=\frac{1}{\sqrt{-g}}\partial_\mu (\sqrt{-g}g^{\mu\nu}\partial_\nu ) \;,
\end{equation}
and $g^{\mu\nu}$ is the inverse of the $4\times 4$ matrix  \begin{equation}\label{acm}
g_{\mu\nu}=\frac{n}{mc}\left(
  \begin{array}{cccc}
   -(c^2-\vec v_0^2) & -v_0^i \\
     -v_0^j & \delta_{ij} \\
  \end{array}\right)\; .
\end{equation}
Here $g$ is the determinant of $g_{\mu \nu}$ and Latin indices range from $1$ to $3$ and are used for spatial components. \par
Eq. (\ref{eqfase}) therefore has the form of the covariant Klein-Gordon equation for a massless scalar field propagating in a fictitious curved spacetime whose line element is
\be \label{lelm}
ds^2=g_{\mu\nu}dx^{\mu} dx^{\nu} =\frac{n}{mc}\left[ -c^2dT^2 + (d\vec x -\vec v_0 dT)(d\vec x -\vec v_0 dT)\right]\ .
\ee
This is the core of the analogy: within the hydrodynamical approximation, the Bogoliubov theory of phase fluctuations in a BEC is equivalent to QFT in curved spacetime for a massless scalar field. Looking at the metric (\ref{acm})
we see that for condensate configurations for which the flow turns supersonic, i.e. admitting a region where $|\vec v_0|>c$, the associated analogue metric describes what gravitational physicists would call a BH in Painvlev\'e-Gullstrand coordinates.\par
On the basis of the analogy one expects this kind of BEC configuration to show a Hawking-like process in the form of emission of correlated pairs of phonons on opposite sides of the horizon. The study of these correlations will be performed using the tools of QFT in curved spacetime for an idealized flow profile.\par
The results will show the validity of this much simpler theoretical scheme to handle a complex system like a supersonic flowing BEC
provided the background quantities vary on scales much larger than $\xi$.

\section{The model}

Our model consists in an infinite elongated (along the $x$ axis) condensate whose transverse size $l_\perp$ is constant
and much smaller than $\xi$.
So the dynamics is frozen in the transverse direction and the problem becomes basically one dimensional.
The flow is stationary and directed along $x$, from right to left, i.e. $\vec v_0=-v_0\hat x$ with $v_0$ a positive constant. The sound speed is adjusted (see below) so that for $x>0$, it is $v_0<c(x)$, while for $x<0$, $v_0>c(x)$. So $x=0$ plays the
role of the horizon and the supersonic ($x<0$) region is the BH region. \par
\begin{figure}[!h]
\begin{center}
\includegraphics[scale=0.6]{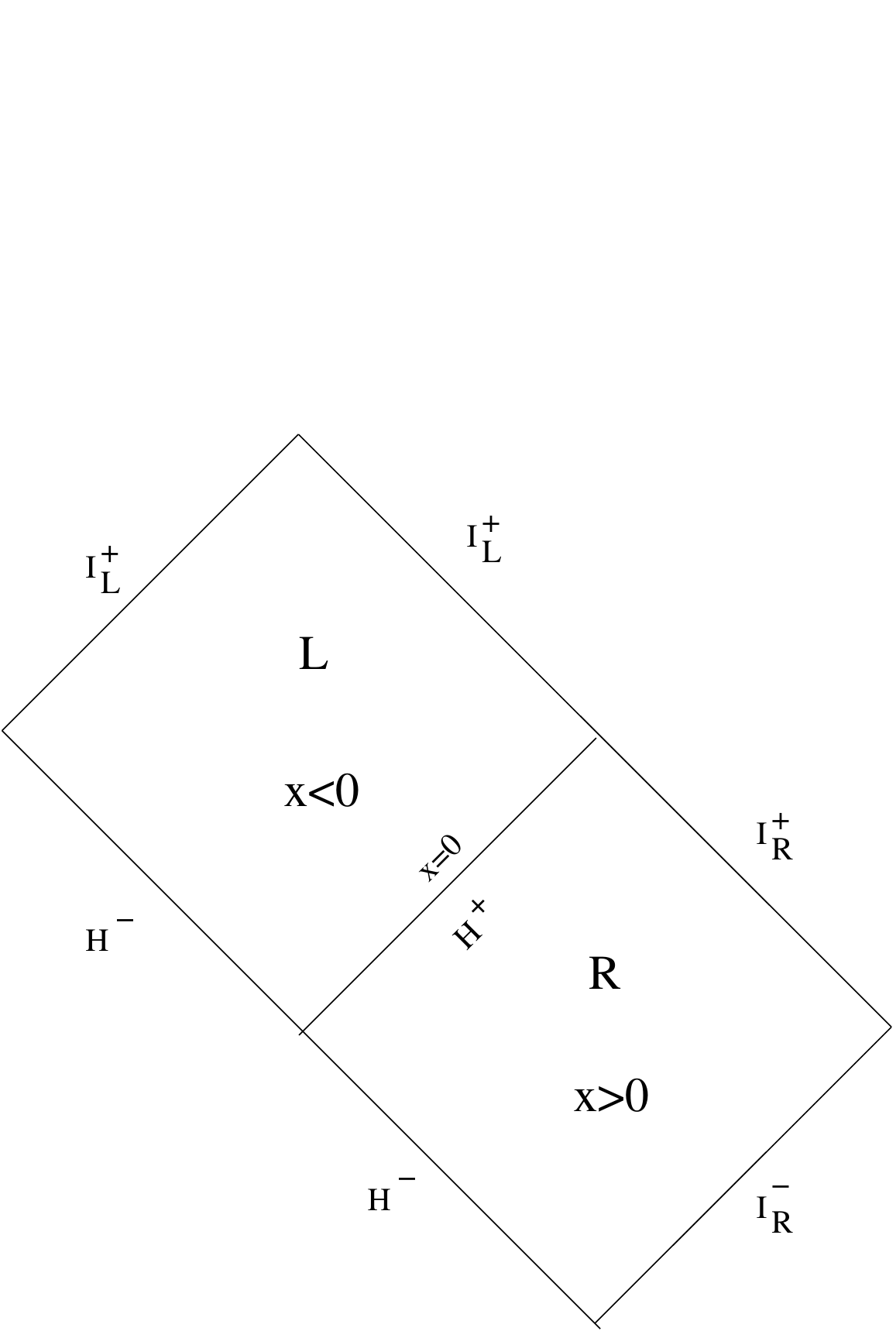}
\end{center}
\caption{Penrose diagram for an acoustic black hole metric.}
\label{fig1}
\end{figure}

We will call the region $x>0$ the $R$ region and the region $x<0$ the $L$ region. A Penrose diagram for the acoustic metric is given in Fig.~\ref{fig1}.
We further assume the density to be constant. Our nontrivial configuration for the sound speed can be obtained by changing the coupling constant $g$ (and hence the speed of sound) along $x$ by means of a spatially varying magnetic field in the vicinity of a Feshbach resonance~\cite{pistri}. \par
Quantization is performed expanding the hermitian quantum operator $\hat\theta_1$ in terms of creation and annihilation operators satisfying usual bosonic commutation rules. Symbolically
\be
\hat\theta_1=\sum_i [ \hat a_if_i + \hat a_i^{\dagger}f_i^*]\ ,
\ee
where the $f_i$ and their complex conjugates are a complete set of solutions to the mode equation
\be \label{boxu}
\hat\Box f_i=0 \;,
\ee
with $\hat\Box$ is given in Eq.~(\ref{box}). The choice of the set of solutions $\{ f_i \}$ selects the vacuum state for the field as
$\hat a_i|0\rangle =0$. The appropriate vacuum to describe Hawking radiation is discussed in the following section. The equation (\ref{boxu}) once expanded has the same form as (\ref{eqfase}) with $\hat \theta_1$ replaced by $f_i$, $\vec v_0=-v_0\hat x$ with $v_0$ and $n$ positive constants.  The modes $f_i$ are assumed to be functions only of $T$ and $x$. Our assumption regarding the scale of the transverse dimensions ($l_\perp \ll \xi$) forbids excitations with transverse momenta.  \par
In order to solve the mode equation a sequence of coordinate transformations, familiar in BH physics, will be performed to simplify it. First we introduce a ``Schwarzschild-like'' time $t$ as follows:
\be t = T -  \int_{x_1}^{x} dy \frac{v_0}{c^2(y) - v_0^2} \label{tTout} \ee
in the $R$ region, while in the $L$ region
\be t = T -  \int_{x_2}^{x} dy \frac{v_0}{c^2(y) - v_0^2} + a \; . \label{tTin} \ee
The constants $x_1,x_2$ and  $a$ are arbitrary.
\par One can then write the mode equation as
\be (\Box^{(2)}  + V) \hat{\theta}_1^{(2)} = 0  \;, \label{boxtheta2} \ee
where $\hat{\theta}_1^{(2)}$ is related to $\hat{\theta}_1$ by
\be  \hat{\theta}_1 = \sqrt{\frac{m c}{n \hbar \ell_\bot^2}} \hat{\theta}_1^{(2)} \;, \label{theta2def} \ee
and
\bea \Box^{(2)} &\equiv & \left[- \frac{c}{c^2-v_0^2} \frac{\partial^2}{\partial t^2} + \frac{c^2-v_0^2}{c} \frac{\partial^2}{\partial x^2} + \frac{dc}{dx} \left(1 + \frac{v_0^2}{c^2} \right) \frac{\partial}{\partial x} \right] \;, \\
     V &\equiv & \frac{1}{2}\frac{d^2 c}{d x^2} \left(1 - \frac{v_0^2}{c^2} \right) - \frac{1}{4c} \left(\frac{d c}{dx}\right)^2 + \frac{5 v_0^2}{4c^3} \left(\frac{d c}{d x}\right)^2  \;. \label{Vdef}
\eea
Note that $\Box^{(2)}$ is the covariant d'Alambertian for the two dimensional metric
\be ds^2 = - \frac{c^2 - v_0^2}{c} d t^2 + \frac{c}{c^2-v_0^2} dx^2 \;. \label{2Dmetric} \ee
 Examination of~\eqref{2Dmetric} shows that unlike the coordinate $T$ which is always timelike, the coordinate $t$ is timelike in the R region and spacelike in the L region.
It approaches the value $-\infty$ on the past horizon, $H_-$, and the value $+\infty$ on the future horizon $H^+$.

It is useful to define in both the $R$ and $L$ regions the ``tortoise'' coordinate $x^*$.  In the $R$ region the definition is
\be  x^* = \int_{x_3}^{x} \frac{c(y)}{c^2(y) - v^2} dy   \;. \label{xstarout} \ee
In the $L$ region the definition is the same but with a different, in general, lower limit which we will call $x_4$.
In the $R$ region $x^*$ ranges from $-\infty$ on the past and future horizons ($H^{\pm}$) to $+ \infty$ in the limit $x \rightarrow + \infty$.
In the $L$ region $x^*$ is again $-\infty$ on the past and future horizons and increases to $+ \infty$ in the limit $x \rightarrow -\infty$.  It thus acts as
a typical time coordinate, as can be seen rewriting the metric as
\be ds^2 = \frac{c^2 - v_0^2}{c} (-dt^2 + dx^{*\,2})  \;. \label{metric-xstar}  \ee
The utility of this tortoise coordinate is that the wave equation for the modes
takes the form
\be  \left(-\frac{\partial^2}{\partial t^2} +  \frac{\partial^2}{\partial x^{* \,2}} + V_{\rm eff}  \right) \hat{\theta}_1^{(2)} = 0 \;, \label{wave-eq-2} \ee
with
\be V_{\rm eff} = \frac{c^2-v_0^2}{c}\, V   \;.  \label{Veff} \ee
With these definitions we can define in both the $R$ and $L$ regions the retarded and advanced null coordinates
respectively
\bea
u &=& t - x^* \;, \nonumber \\
v &=& t + x^* \;.
\eea
Because the mode functions propagate across the future horizon $I^+$, it is useful to have $v$ be a continuous variable across that horizon.  As shown in Sec.~VIII
one way to accomplish this is to
choose $x_1=x_2$, $x_3=x_4$, and then find the appropriate value for $a$.
\par For later use we introduce another set of retarded and advanced coordinates called Kruskal coordinates which are defined as
\bes
\bea  U_K &=& - e^{- \kappa u}/\kappa \;, \\
      V_K &=&  e^{\kappa v}/\kappa \;,
\eea \ees
in $R$ region, while
\bes \bea      U_K &=&  e^{-\kappa u} /\kappa \;,  \\
               V_K &=& e^{\kappa v}/\kappa \;,
\eea \ees
in $L$ region. Therefore $U_K=0$ on the future horizon $H^+$ and $V_K = 0$ on the past horizon $H^-$. The parameter $\kappa$ is the surface gravity of the horizon, in our acoustic setting it is
\be \kappa = \frac{dc}{dx}|_{x=0}\ .\ee
In Ref.~\cite{paper1}, the term containing $V_{eff}$ in the mode equation~(\ref{wave-eq-2}) was neglected with the result that the mode solutions can be written in terms of the simple plane waves $(e^{-iwu}$ and $\ e^{-iwv})$ that propagate freely without being scattered.  Here the full equation (with $V_{eff}$ included) will be solved (numerically) to also take into account
the backscattering which in the 4-D case is caused by the effective curved geometry (i.e. the inhomogeneities of the BEC medium) and in the 2-D case is caused by the effective potential.

\section{The Unruh State}

To proceed further one has to select the quantum state for the field $\hat\phi$ encoding the formation of the acoustic BH which is the process which triggers Hawking radiation.\par
However if one limits our analysis to sufficiently late times after the BH formation, the features of the emitted Hawking quanta are completely independent on the details of the formation process, being determined only by the properties of the resulting stationary horizon, more specifically its surface gravity $\kappa$. So one can avoid discussing the complicated physics underlying the dynamics of the formation process and simply impose on the past horizon $H^-$ of our stationary BH metric the appropriate, but universal, boundary conditions which mimic the horizon formation process. \par
This is the spirit of the so called Unruh state~\cite{unruh}. Technically Hawking radiation is superimposed on our stationary background by requiring the retarded modes originating from the past horizon $H^-$ to be positive frequency with respect to the Kruskal coordinate $U_K$ which is an affine parameter along $H^-$.

\begin{figure}[!h]
\begin{center}
\includegraphics[scale=0.6]{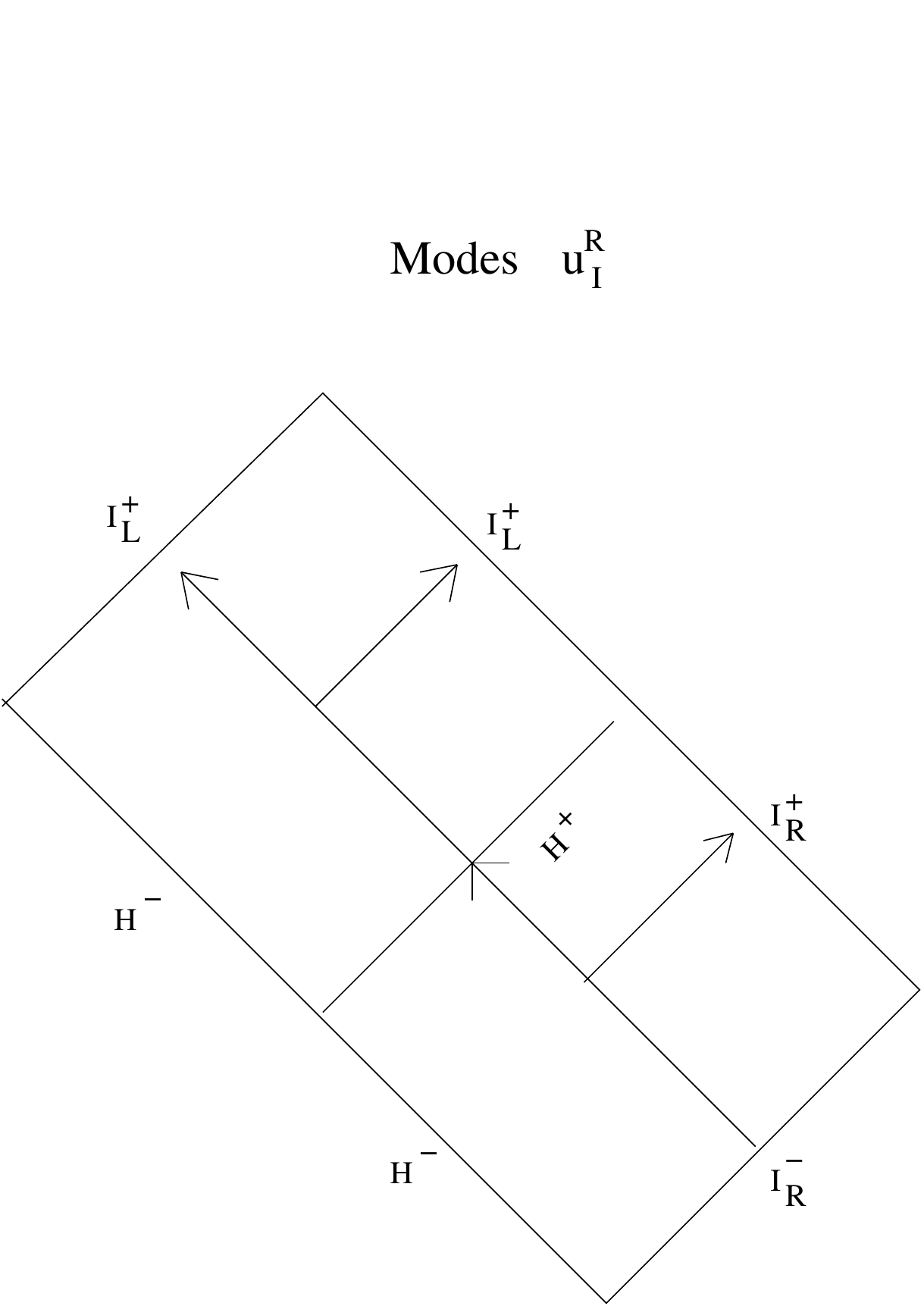}
\end{center}
\caption{Incoming modes $u_I^R$ from $I_R^-$.}
\label{fig2}
\end{figure}

For this state there is a flux of radiation which comes through
past horizon, $H^-$ while at past null infinity, i.e. $I^-_R$, there is no radiation.  To obtain this behavior the scalar field $\hat{\theta}_1^{(2)}$ is expanded in terms of two sets of modes so that
\bea \hat{\theta}_1^{(2)} &=& \int_0^\infty d \omega_K  \left[ a_K(\omega) u^K_H(\omega,x) + a_K^\dagger(\omega) u^{K\,*}_H(\omega,x) \right] \nonumber \\
 & & + \int_0^\infty d \omega \left[ b(\omega) u^{\rm out}_I(\omega,x) + b^\dagger(\omega) u^{{\rm out}\,*}_I(\omega,x) \right]  \;.  \label{phi-modes}
\eea
Here both sets of creation and annihilation operators obey the usual commutation relations and both sets of modes are solutions to the mode equation
\be  \left(-\frac{\partial^2}{\partial t^2} +  \frac{\partial^2}{\partial x^{* \,2}} + V_{\rm eff}  \right) u(\omega,x) = 0 \;. \label{mode-eq} \ee
The modes $u^{R}_I$ originate at $I^-_R$.  They are positive frequency
with respect to the time coordinate $t$ and on $I^-_R$ have the form $u^{R}_I \sim e^{- i \omega v}$. Because of the potential term in Eq.~(\ref{mode-eq}) they are partly transmitted towards $I_L^+$ and partially reflected to $I^+_R$ and $I^+_L$, see Fig.~\ref{fig2}.

\begin{figure}[!h]
\begin{center}
\includegraphics[scale=0.6]{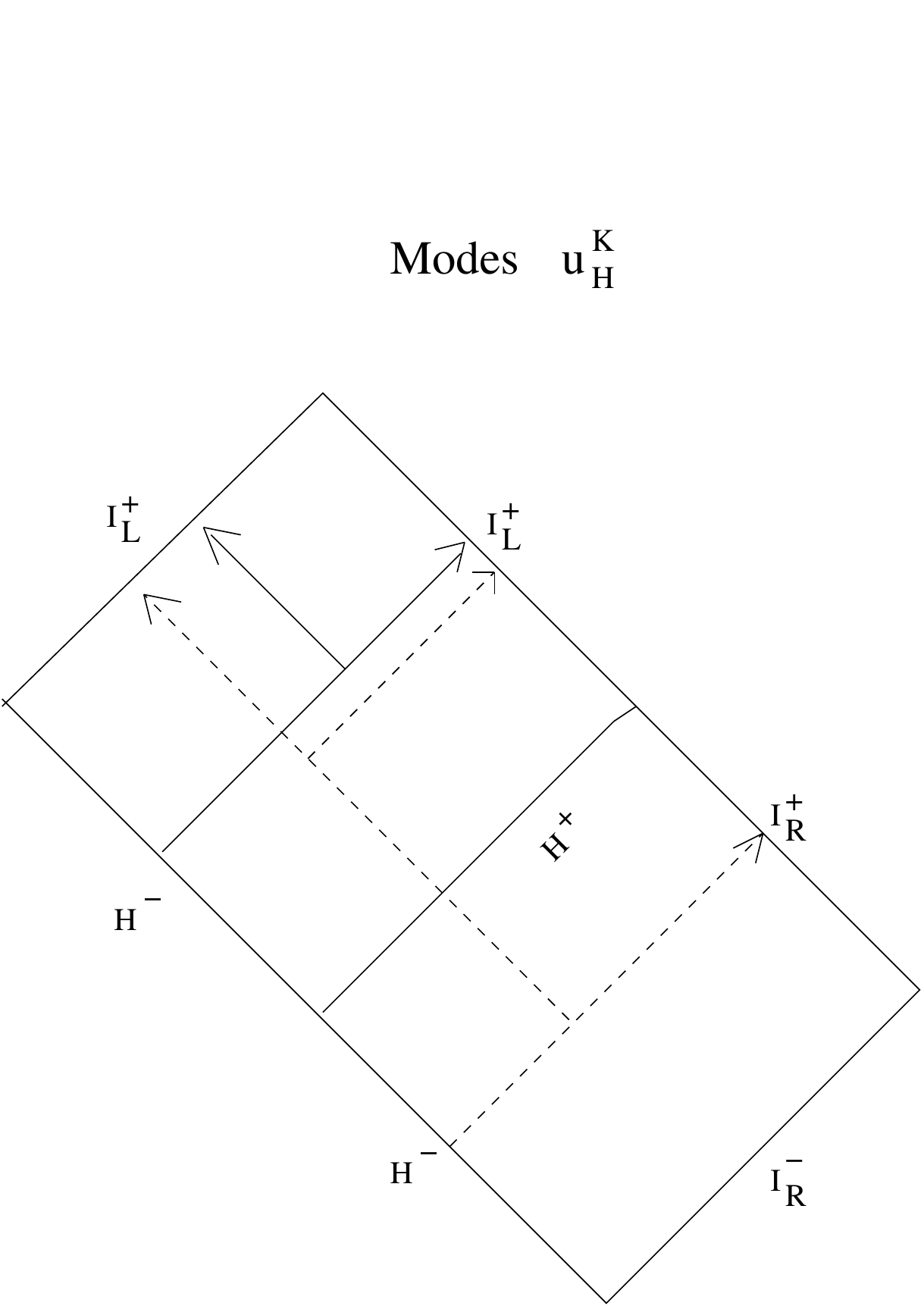}
\end{center}
\caption{Outgoing modes $u_H^K$ from $H^-$.}
\label{fig3}
\end{figure}

The modes $u^{K}_H$ which come
through $H^-$ are, on $H^-$, positive frequency with respect to the Kruskal null coordinate $U_K$.  There they have the form
$u^{K}_H \sim e^{-i \omega_K U_K}$. They are partially transmitted to  $I^+_R$ or $I^+_L$ and partially reflected to
$I^+_L$, see Fig.~\ref{fig3}.

Note that these behaviors on $I^-_R$ and $H^-$ define the state because it is always possible to fix the mode function
and its first derivative in any way that one wishes on a Cauchy surface.  Different ways of fixing these initial conditions lead to different states for the field.

The modes are normalized using the scalar product~\cite{b-d-book}
\be  (\phi_1,\phi_2) = -i \int_\Sigma \, \phi_1(x) \stackrel{\leftrightarrow}{\partial}_\mu \phi^*_2(x) \sqrt{-g_\Sigma(x)} n^\mu d\Sigma  \;, \label{scalar-product}
\ee
with $\Sigma$ a Cauchy surface and $n^\mu$ a future directed unit vector which is perpendicular to $\Sigma$.
The normalization condition is
\be (u(\omega,x),u(\omega',x)) = \delta(\omega-\omega')  \;. \label{unorm} \ee

Normalization of the $u_{I}$ modes is done on the surface $I^-_R$.  Technically one needs a Cauchy surface.  A full Cauchy surface for the $R$
region would consist of $I^-_R$ along with the part of $H^-$ which bounds this region. Then one could add to this a spacelike or null surface covering
the rest of the spacetime.  What is chosen is unimportant because the modes which originate on $I^-_R$ are zero on the past horizon and along the above mentioned surface.
On the surface $I^-_R$, the normalization condition is
\be (u^{\rm out}_I(\omega,x), u^{\rm out}_I(\omega',x)) = -i \int_{-\infty}^\infty \, d v \,  u^{\rm out}_I(\omega,x) \stackrel{\leftrightarrow}{\partial}_v  u_I^{{\rm out}\,*}(\omega',x)   = \delta(\omega-\omega') \;. \ee
The result is that on $I^-_R$,
\be  u^{\rm out}_I(\omega,x) = \frac{e^{-i \omega v}}{\sqrt{4 \pi \omega}}  \;. \label{uI-norm}  \ee

Normalization of the the $u^K_H$ modes is done along $H^-$ which is a Cauchy surface.  The normalization condition is
\be (u^{K}_H(\omega_K,x), u^{K}_H(\omega_K',x)) = -i \int_{-\infty}^\infty \, d U_K  \, u^{K}_H(\omega_K,x) \stackrel{\leftrightarrow}{\partial}_{U_K}  u_H^{K\,*}(\omega_K',x)   = \delta(\omega_K-\omega_K') \;. \ee
The result on $H^-$ is
\be u^K_H(\omega,x) = \frac{e^{-i \omega_K U_K}}{\sqrt{4 \pi \omega_K}}  \;. \label{uK-norm}  \ee

Although one can define initial data for the $u^K_H$ modes on $H^-$, it is difficult to evolve these modes because in the Kruskal coordinates $U_K$ and $V_K$ the mode equation is not separable and in the $t$ and $x$ coordinates it is not easy to express the initial conditions for these modes.  The mode solutions discussed above that begin on $I^-_R$ are zero on $H^-$.  However, there exists a set of solutions to the mode equation~\eqref{mode-eq} which are positive frequency with respect to $t$ and which pass through the part of $H^-$ which intersects the $R$ region.  On that horizon they have the behavior $u^{R}_{H} \sim e^{-i \omega u}$.

\begin{figure}[!h]
\begin{center}
\includegraphics[scale=0.6]{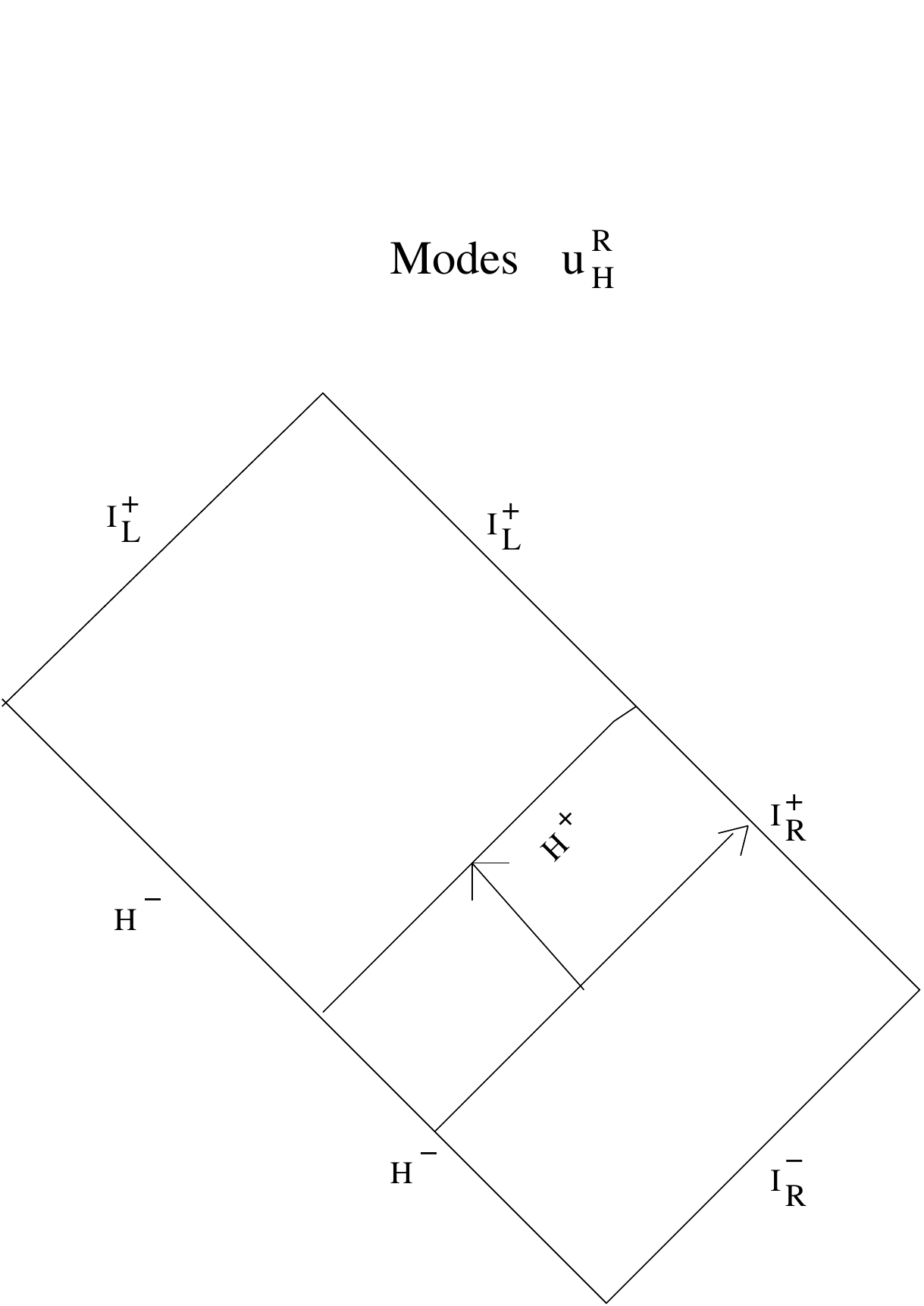}
\end{center}
\caption{Ougoing modes $u_H^R$, defined in the exterior region ($x>0$).}
\label{fig4}
\end{figure}

These solutions are zero on both $I^-_R$ and the part of $H^-$ which borders the {\it in} region.  They are normalized on $H^-$ using the condition
\be (u^{R}_H(\omega,x), u^{R}_H(\omega',x)) = -i \int_{-\infty}^\infty \, d u  \, u^{R}_H(\omega,x) \stackrel{\leftrightarrow}{\partial}_u  u^{{R}\,*}_H(\omega',x)   = \delta(\omega-\omega') \;, \label{H-norm} \ee
with the result that on $H^-$,
\be  u^{R}_{H}(\omega,x) = \frac{e^{-i \omega u}}{\sqrt{4 \pi \omega}}  \;, \label{uH-out-norm}  \ee
see Fig.~\ref{fig4}.

\begin{figure}[!h]
\begin{center}
\includegraphics[scale=0.6]{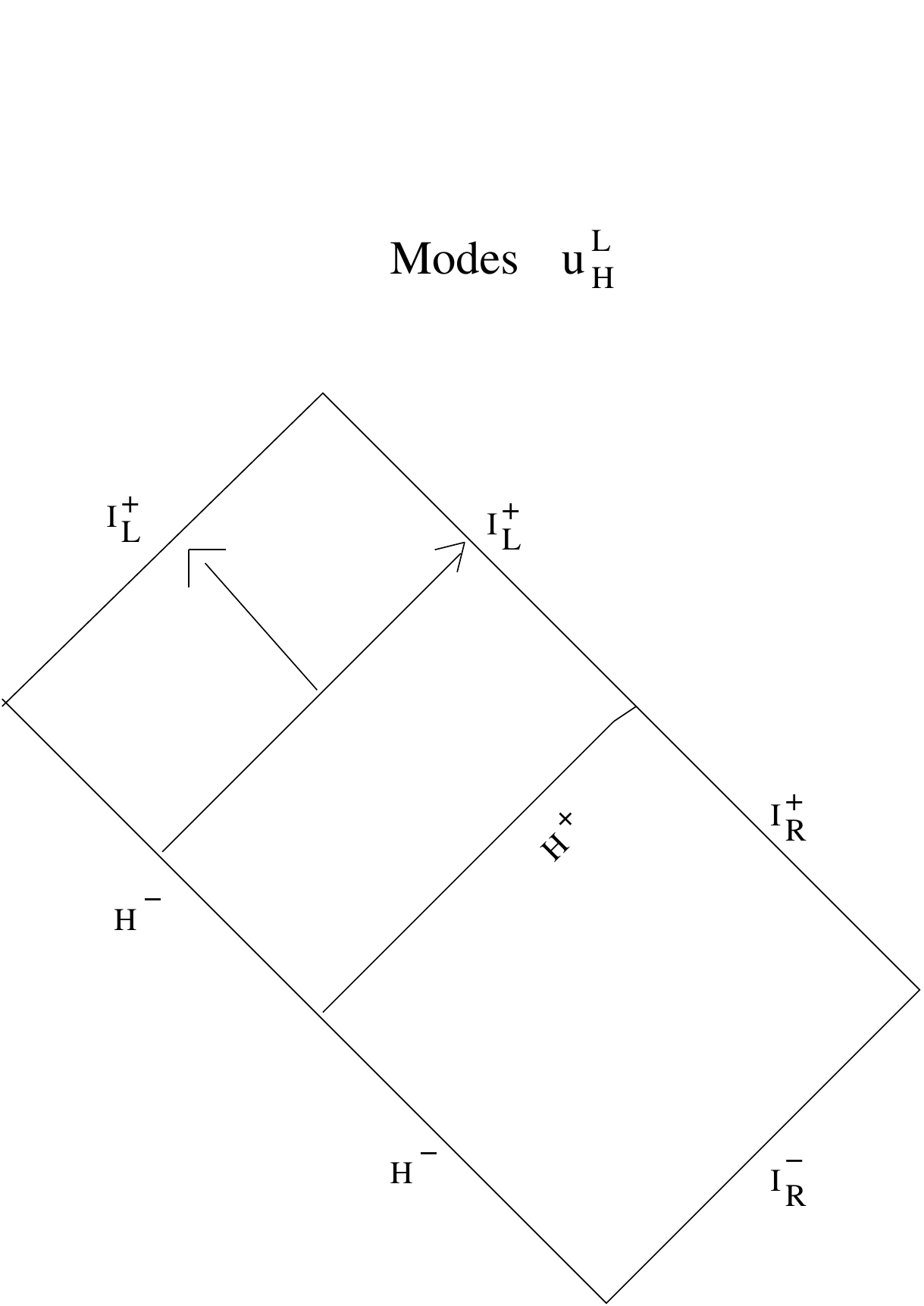}
\end{center}
\caption{Outgoing modes $u_H^L$, defined in the interior region ($x<0$).}
\label{fig5}
\end{figure}

For the part of $H^-$ which serves as a boundary for the $L$ region one has a complete set of modes which on $H^-$ are positive frequency with respect to the time coordinate $x^*$.  They are zero on the part of $H^-$ which borders the
$R$ region.  On the other part of $H^-$ they have the form $u^{L}_{H^-} \sim e^{i \omega u}$ and are normalized using the condition~\eqref{H-norm} with the result that on the part of $H^-$ which borders the {\it in} region,
\be  u^{L}_{H}(\omega,x) = \frac{e^{i \omega u}}{\sqrt{4 \pi \omega}}  \;, \label{uH-in-norm}  \ee
see Fig.~\ref{fig5}.

All that remains is to express the modes $u_H^K$ in terms of the modes $u^{L}_{H}$ and $u^{R}_{H}$.  This is done through Bogolubov transformations whose coefficients can be computed using the scalar product evaluated on the past horizon $H^-$.  We can write
\bea u^K_{H}(\omega_K,x) &=&  \int_0^\infty d \omega \, \left[ \alpha^{L}_{\omega_K \omega} u^{L}_{H}(\omega,x) + \beta^{L}_{\omega_K \omega} u^{L \, *}_{H}(\omega,x)  \right]  \nonumber \\
  \,&+&\, \int_0^\infty d \omega \, \left[ \alpha^{R}_{\omega_K \omega} u^{R}_{H}(\omega,x) + \beta^{R}_{\omega_K \omega} u^{R \, *}_{H}(\omega,x)  \right] \;. \label{Bog1}
\eea
The coefficients can be obtained by using the scalar product evaluated on $H^-$.  Recalling that $u^{R}_{H^-} = 0$ on the part of $H^-$ which borders the $L$ region and $u^{L}_{H^-} = 0$ on the part of $H^-$ which borders the $R$ region we have
\begin{subequations}
\bea
\alpha^{R}_{\omega_K \, \omega} &=& (u^K_{H}(\omega_K,x), u^{R}_{H}(\omega,x))  \;, \label{alpha-out-1} \\
      \beta^{R}_{\omega_K \, \omega} &=& -(u^K_{H}(\omega_K,x), u^{L \,*}_{H}(\omega,x))  \;, \label{beta-out-1} \\
\alpha^{L}_{\omega_K \, \omega} &=& (u^K_{H}(\omega_K,x), u^{L}_{H}(\omega,x))  \;, \label{alpha-in-1} \\
      \beta^{L}_{\omega_K  \, \omega} &=& - (u^K_{H}(\omega_K,x), u^{L \,*}_{H}(\omega,x))  \ .  \label{beta-in-1}
\eea
\end{subequations}
Using the expressions~\eqref{uK-norm} and~\eqref{uH-out-norm} and evaluating the scalar product on $H^-$ one finds
\begin{subequations}
\bea \alpha^{R}_{\omega_K \, \omega} &=& \frac{-i}{4 \pi \sqrt{\omega_K \, \omega}} \int_{-\infty}^0 d U_K \exp \left[-i \omega_K U_K - i \frac{\omega}{\kappa} \log (-U_K) \right] \, \left[ \frac{i \omega}{\kappa (-U_K)} + i \omega_K \right] \ . \ \ \ \ \ \
\eea
Making the change of variables $Z = - U_K$ allows one to compute the integral in terms of Gamma functions with the result that
\be \alpha^{R}_{\omega_K \, \omega}  = \frac{1}{2 \pi \kappa} \sqrt{\frac{\omega}{\omega_K}} \, \Gamma(-i \omega/\kappa) (-i \omega_K)^{i  \omega/\kappa}  \;.  \label{alpha-out} \ee
In a similar way one finds
\bea \beta^{R}_{\omega_K \, \omega} &=& \frac{i} {4 \pi \sqrt{\omega_K \, \omega}} \int_{-\infty}^0 d U_K \exp \left[-i \omega_K U_K + i \frac{\omega}{\kappa} \log (-U_K) \right] \, \left[ \frac{-i \omega}{\kappa (-U_K)} + i \omega_K \right] \nonumber \\
 &=& \frac{1}{2 \pi \kappa} \sqrt{\frac{\omega}{\omega_K}} \, \Gamma(i \omega/\kappa) (-i \omega_K)^{-i  \omega/\kappa}  \;,  \label{beta-out} \\
 \alpha^{L}_{\omega_K \, \omega} &=& -\frac{i}{4 \pi \sqrt{\omega_K \, \omega}} \int_0^\infty d U_K \exp \left[-i \omega_K U_K + i \frac{\omega}{\kappa} \log (U_K) \right] \, \left[ \frac{i \omega}{\kappa U_K} + i \omega_K \right] \nonumber \\
 &=& \frac{1}{2 \pi \kappa} \sqrt{\frac{\omega}{\omega_K}} \, \Gamma(i \omega/\kappa) (i \omega_K)^{-i  \omega/\kappa}  \;,  \label{alpha-in} \\
 \beta^{L}_{\omega_K \, \omega} &=& \frac{i}{4 \pi \sqrt{\omega_K \, \omega}} \int_0^\infty d U_K \exp \left[-i \omega_K U_K - i \frac{\omega}{\kappa} \log (U_K) \right] \, \left[ \frac{-i \omega}{\kappa U_K} + i \omega_K \right] \nonumber \\
 &=& \frac{1}{2 \pi \kappa} \sqrt{\frac{\omega}{\omega_K}} \, \Gamma(-i \omega/\kappa) (i \omega_K)^{i  \omega/\kappa}  \;.  \label{alpha-in}
\eea
\label{bogolubov}
\end{subequations}

\section{Density correlations}

\begin{figure}[!h]
\begin{center}
\includegraphics[scale=0.32]{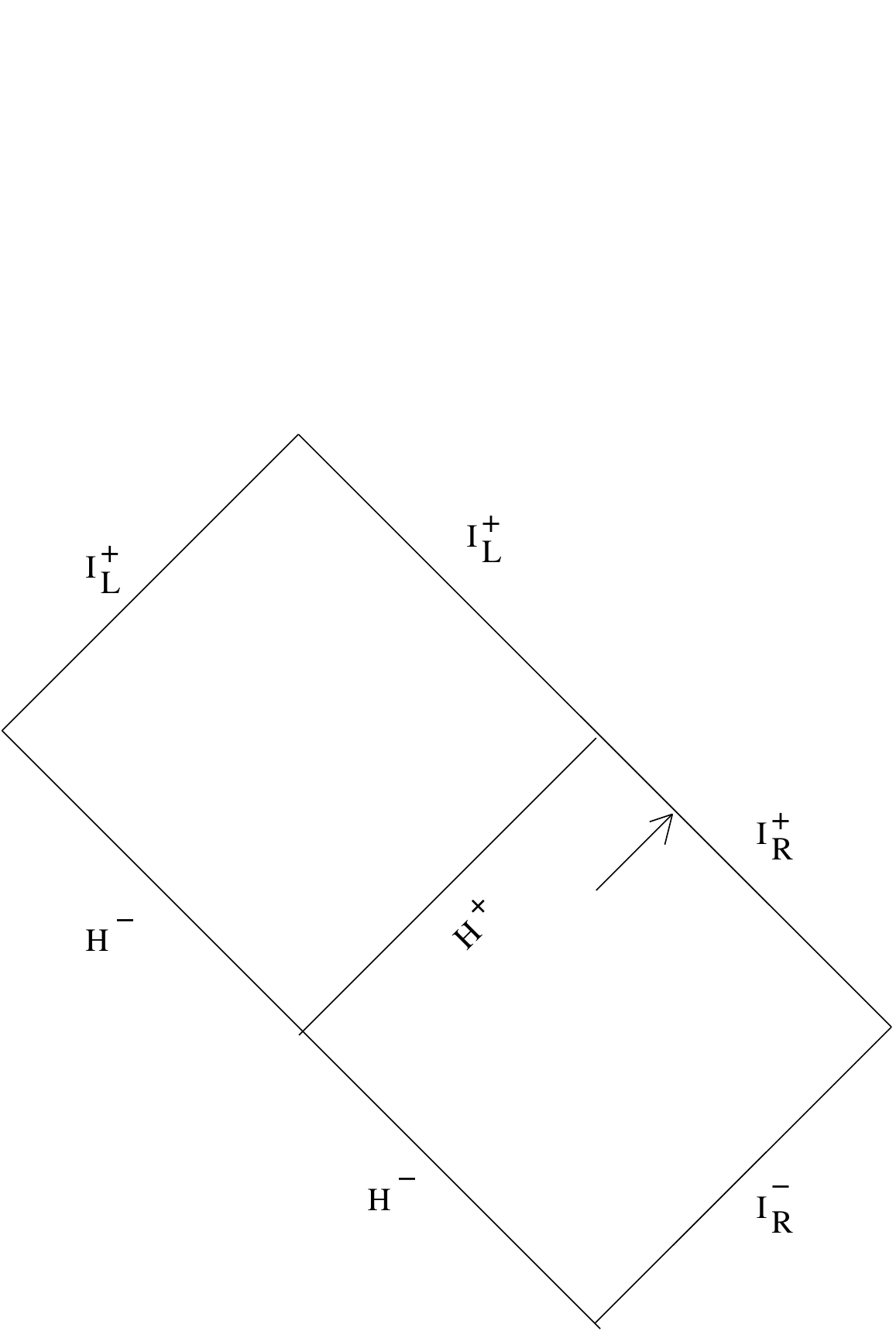}
\end{center}
\caption{Phonons (Hawking quanta) created in the exterior (subsonic) region. }
\label{fig6}
\end{figure}

\begin{figure}[!h]
\begin{center}
\includegraphics[scale=0.32]{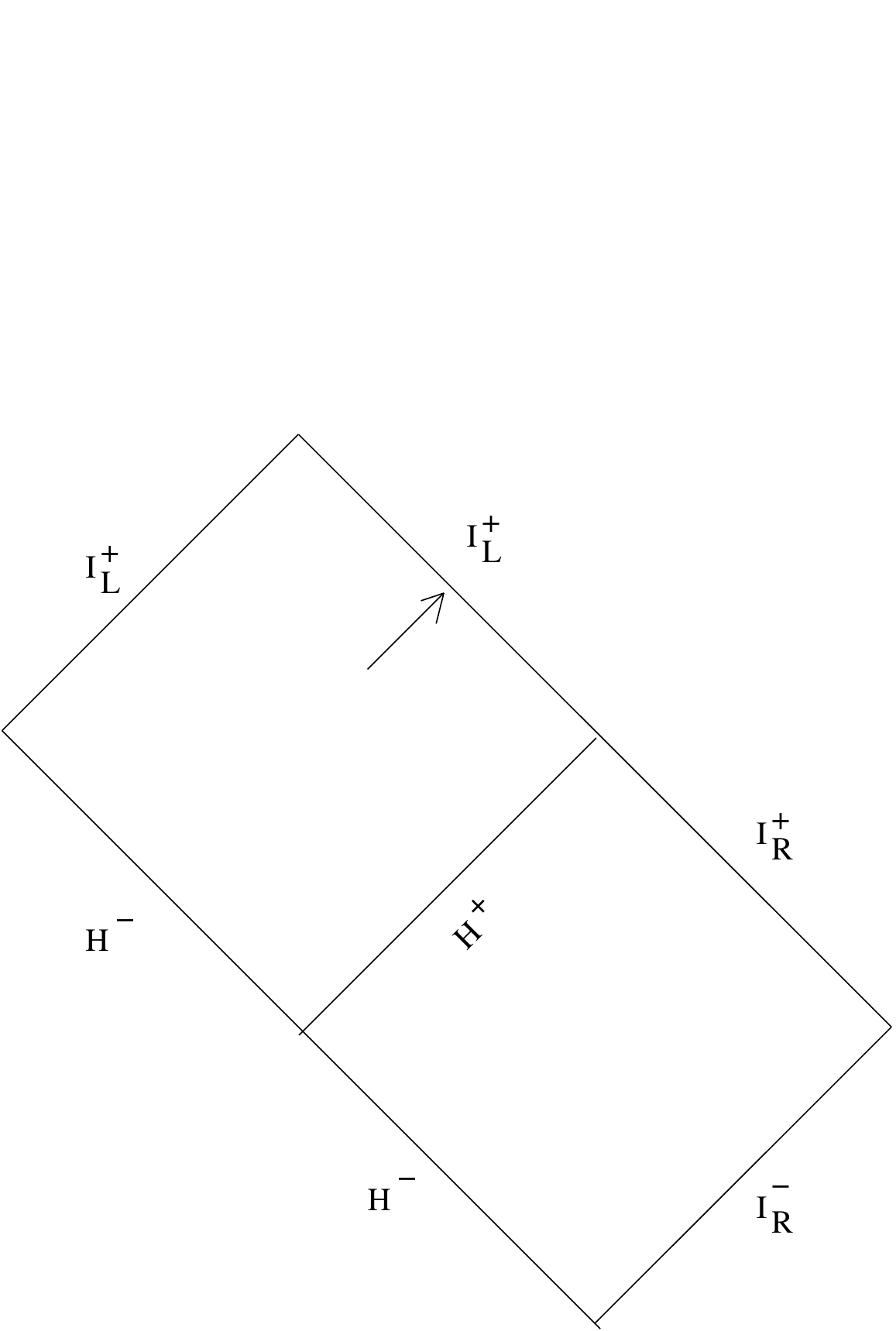}
\end{center}
\caption{Negative frequency phonons (partners) created in the interior (supersonic) region.}
\label{fig7}
\end{figure}

\begin{figure}[!h]
\begin{center}
\includegraphics[scale=0.32]{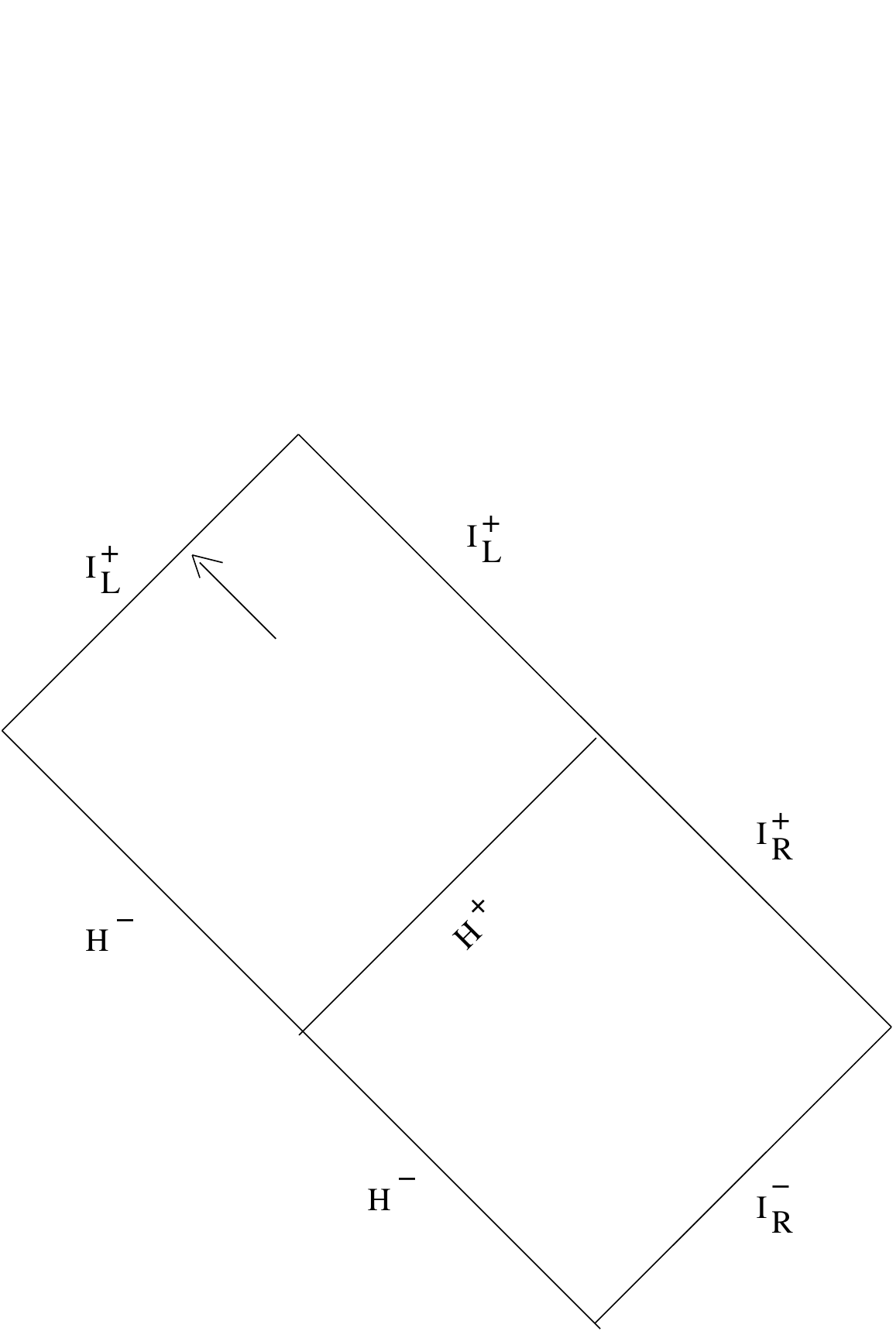}
\end{center}
\caption{Positive frequency phonons created in the interior (subsonic) region.}
\label{fig8}
\end{figure}

Correlations are a manifestation of the entanglement existing between created particles and their partners~\cite{paper1,paper2}, see also~\cite{giovanazzi,max1,max2}.  As discussed in Sec.~IV, the Unruh vacuum we have constructed, i.e. the state annihilated by the operators $\hat a_K$ and $\hat b$ in Eq.~(\ref{phi-modes}) describes Hawking radiation at late times. The created quanta appear at late times as the three types of modes which are depicted in Figs.~\ref{fig6},~\ref{fig7}, and~\ref{fig8}.

The first mode (Fig.~\ref{fig6}) at late times for $x\to \infty$ has form $\frac{e^{-iwu}}{\sqrt{w}}$ and describes phonons propagating towards $x\to +\infty$ with velocity $V=(c_+-v_0)>0$ where $c_+\equiv c(x=+\infty)$.
The second mode (Fig.~\ref{fig7}) at late times for $x\to -\infty$ has the form $\frac{e^{iwu}}{\sqrt{w}}$ and describes phonons propagating inside the BH towards $x\to -\infty$ with velocity $V=(c_--v_0)<0$ where $c_-\equiv c(x=-\infty)$.
These are the so called partners (negative Killing frequency).  The last one  (Fig.~\ref{fig8}) also describes modes propagating inside the BH, this time with velocity $V=-(c_- +v_0)<0$; their asymptotic ($x\to -\infty$) form is
 $\frac{e^{-iwv}}{\sqrt{w}}$ (positive frequency).

\par
\begin{figure}[!h]
\begin{center}
\includegraphics[scale=0.32]{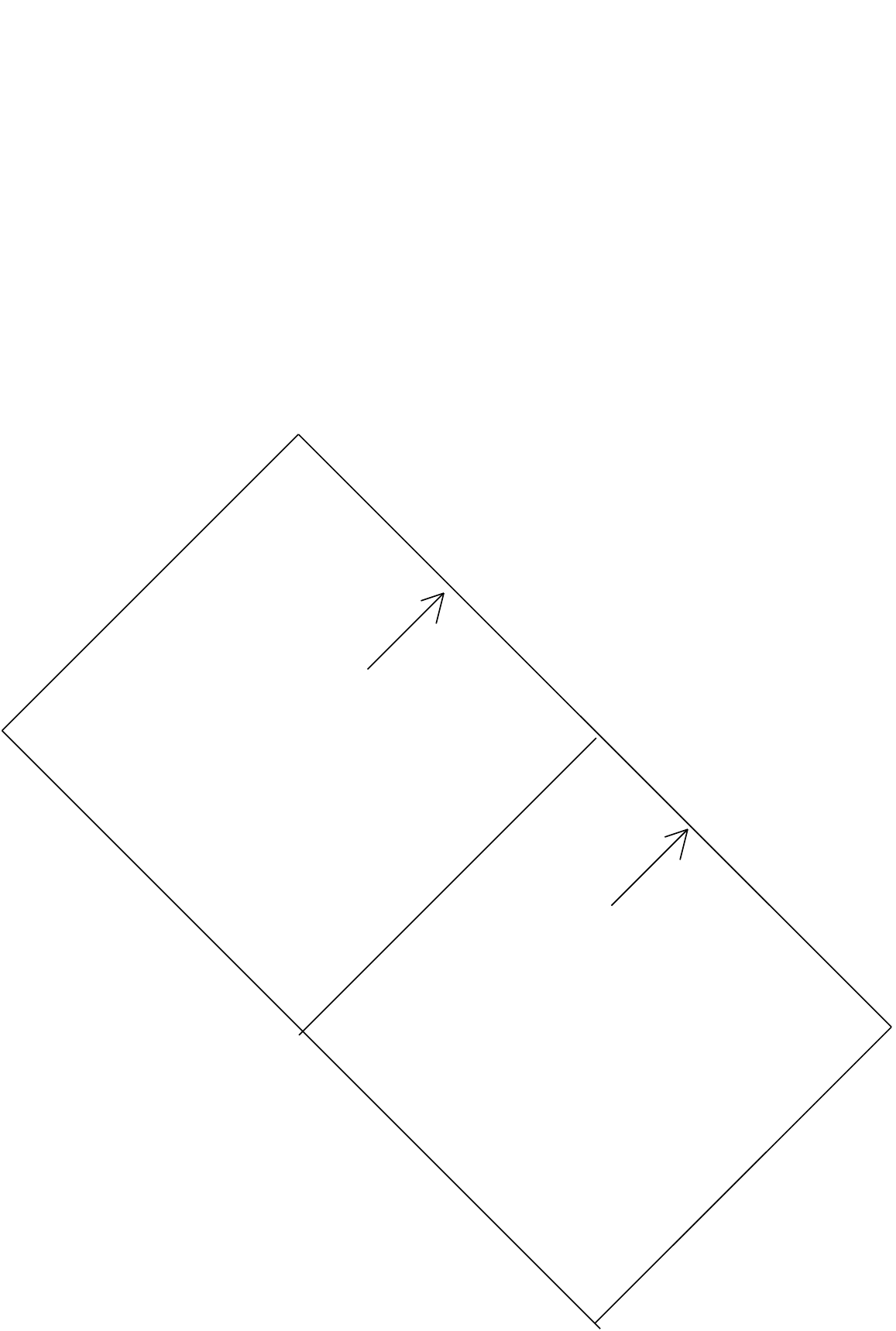}
\end{center}
\caption{Left-Right correlations between phonons in Figs.~\ref{fig6}, ~\ref{fig7}. }
\label{fig9}
\end{figure}

\begin{figure}[!h]
\begin{center}
\includegraphics[scale=0.32]{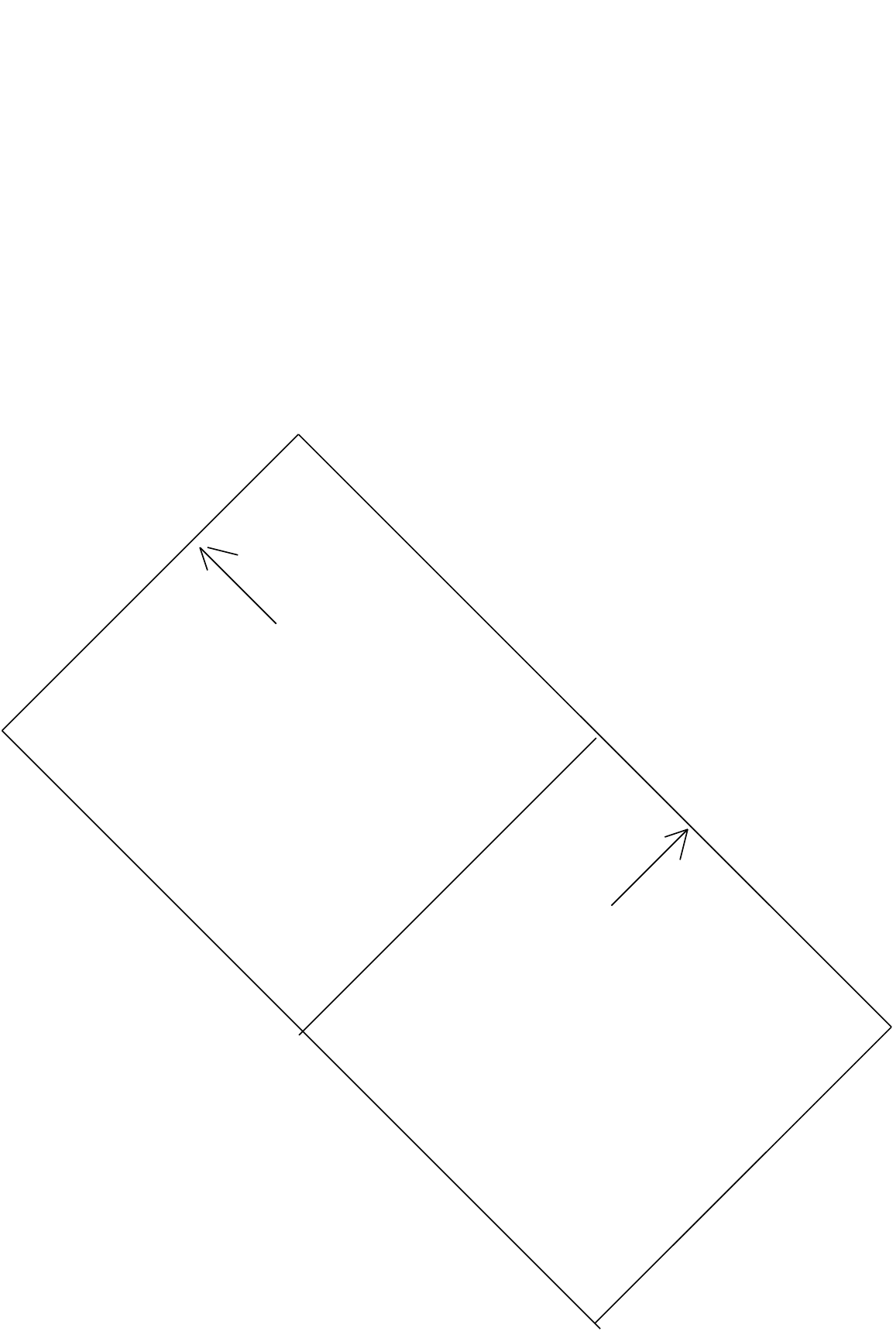}
\end{center}
\caption{ Left-Right correlations between phonons in Figs.~\ref{fig6}, ~\ref{fig8}. }
\label{fig10}
\end{figure}

\begin{figure}[!h]
\begin{center}
\includegraphics[scale=0.32]{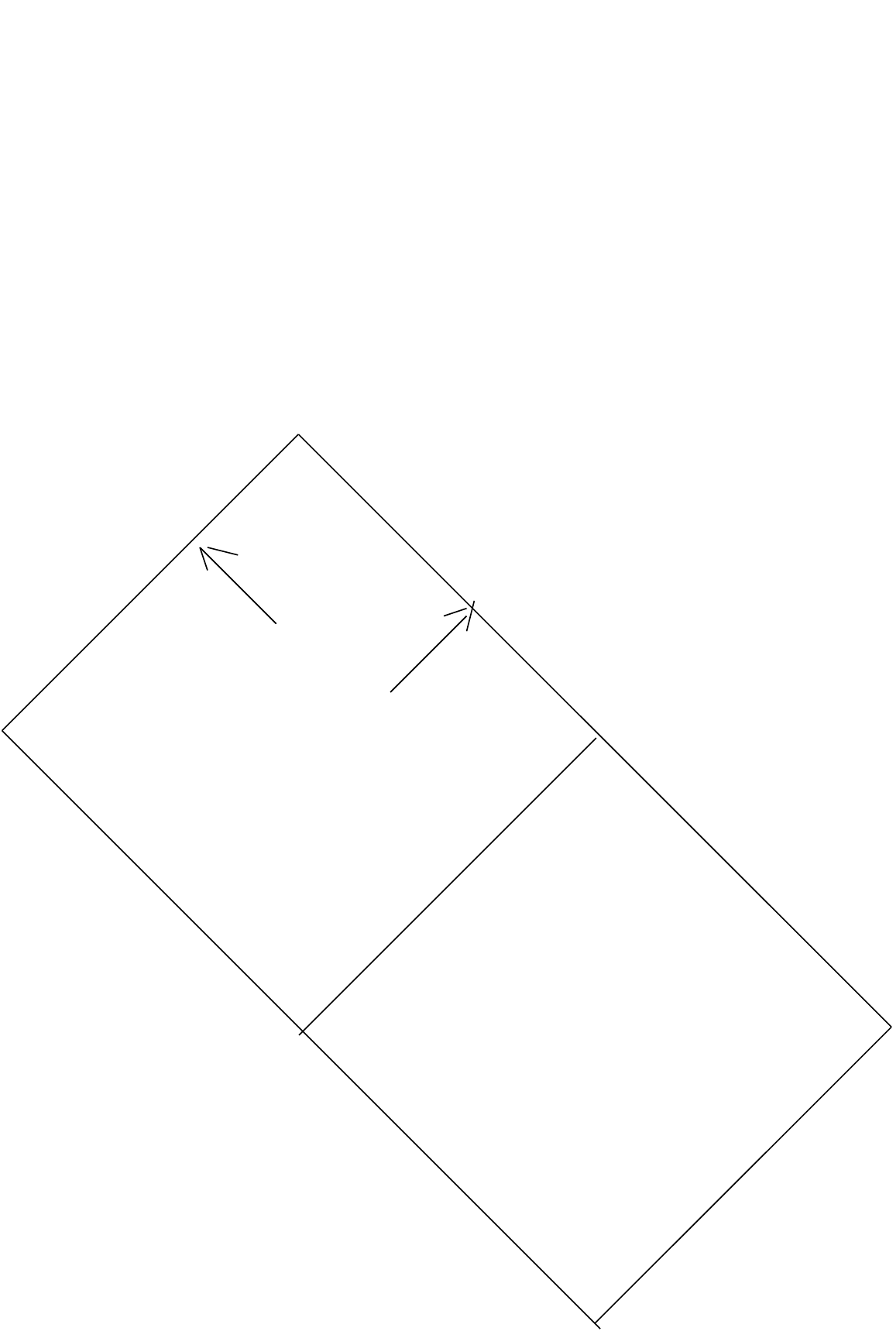}
\end{center}
\caption{Left-Left correlations between phonons in Figs.~\ref{fig7}, ~\ref{fig8}.}
\label{fig11}
\end{figure}

 Two-point correlations related to Hawking radiation are between these modes. There are therefore three kinds of relevant correlations. The first is  $L-R$ and correlates quanta between the modes in Fig.~\ref{fig6} and Fig.~\ref{fig7}, i.e. modes on opposite sides of the horizon, see Fig.~\ref{fig9}.
Associated with this one expects to observe a peak for the correlations in the $(x,x')$ plane along the line
\be \frac{x'}{c_--v_0}=\frac{x}{c_+ -v_0}\ \ee
with $x'<0$ and $x>0$.
There is another correlation of the $L-R$ form which correlates quanta between the modes in Fig.~\ref{fig6} and Fig.~\ref{fig8},
see Fig.~\ref{fig10}.
The peak is along
\be \frac{x'}{-(c_-+v_0)}=\frac{x}{c_+-v_0}\ .\ee
Finally there is a $L-L$ correlation (both modes inside the horizon) of the modes in Fig. ~\ref{fig7} and Fig. ~\ref{fig8},
see Fig. ~\ref{fig11}.
The peak this time is along
\be \frac{x'}{-(c_-+v_0)}=\frac{x}{c_+-v_0} \ee
where both $x,x'<0$.

The equal time density density correlation function, which is the basic object of our investigation, is formally defined as
\be  G_2(T,x;T',x')=\lim_{T\to T'} \langle \hat n_1(T,x) \hat n_1(T',x')\rangle \ .\ee
Writing $\hat n_1$ in terms of the phase operator Eq. (\ref{bhu}) one can relate $G_2$ to the symmetric two-point function of the field $\hat \Theta_1^{(2)}$
\begin{subequations}
\bea  G_2(T,x;T',x') &=& \frac{\hbar n}{2 m \ell_{\perp} c^2(x) c^2(x')} \lim_{T' \rightarrow T} D \sqrt{c(x) c(x')} \, \langle \{ \hat{\theta}_1^{(2)}(t,x), \hat{\theta}_1^{(2)}(t',x') \} \rangle \;,  \\
D & \equiv & \partial_T  \partial_{T'} - v_0 \partial_x \partial_{T'}  - v_0 \partial_T \partial_{x'} + v_0^2 \partial_x \partial_{x'} \;.
\eea
\label{G2formula}
\end{subequations}
where $\{ , \}$ is the anticommutator.
Here one should use the definition~\eqref{tTout} for $t$ in the $R$ region and the comparable definition in the $L$ region.

The two point function is computed using Eqs.~\eqref{phi-modes},   with the result that
\bes
\bea & & \langle \{ \hat{\theta}_1^{(2)}(t,x), \hat{\theta}_1^{(2)}(t',x') \} \rangle =  I  + J \;, \label{IplusJ} \\
 & &  I = \int_0^\infty d \omega_K \, \left[ u^K_H(\omega_K,t,x) \, u^{K \, *}_H(\omega_K,t',x')
   +  u^{K \, *}_H(\omega_K,t,x) \, u^{K}_H(\omega_K,t',x') \right] \;, \label{Idef} \\
 & &  J = \int_0^\infty d \omega \, \left[ u^{R}_I(\omega,t,x) \, u^{{R} \, *}_I(\omega,t',x')
   +  u^{{R} \, *}_I(\omega,t,x) \, u^{R}_I(\omega,t',x') \right] \;. \label{Jdef}
\eea
\label{two-point-1}
\ees
Substituting Eq.~\eqref{Bog1} into Eq.~\eqref{Idef} and using Eqs.~\eqref{bogolubov} one finds $16$ integrals of the form
\begin{eqnarray}
&& \frac{1}{4 \pi^2 \kappa^2}
\int_0^\infty d \omega_K \int_0^\infty d \omega \int_0^\infty d \omega'  \frac{\sqrt{\omega \omega'}}{\omega_K} \Gamma(\pm i \omega/\kappa) \Gamma(\pm i \omega'/\kappa) \nonumber \\ && (\pm i \omega_K)^{\pm i  \omega/\kappa}  (\pm i \omega_K)^{\pm i \omega'/\kappa} u_1(\omega,t,x) u_2(\omega',t',x') \ \ \ \ \ \ \ \ \end{eqnarray}
with all combinations of $\pm$ and $u_1$ and $u_2$ being various combinations of $u^{L}_H$, $u^{R}_H$, and their complex conjugates.  The integral over $\omega_K$ can be computed.  With the change of variable $z = (\ln \omega_K)/\kappa$ the result is proportional to either $\delta(\omega-\omega')$ or  $\delta(\omega+\omega')$.  Given the limits of integration terms containing $\delta(\omega+\omega')$ integrate to zero.  Using the relation
\be \Gamma\left(\frac{i \omega}{\kappa}\right) \, \Gamma\left(\frac{-i \omega}{\kappa}\right) = \frac{\pi \kappa}{\omega \sinh\left(\frac{\pi \omega}{\kappa}\right)}  \;, \ee
one finds that
\bea  I &=& \int_0^\infty d \omega \frac{1}{\sinh\left(\frac{\pi \omega}{\kappa}\right)} \left\{ u^{L}_H(\omega,t,x) \, u^{R}_H(\omega,t',x') +
   u^{L *}_H(\omega,t,x) \, u^{R *}_H(\omega,t',x') \right. \nonumber \\
   & & \;\;\; \left. + u^{R}_H(\omega,t,x) \, u^{L}_H(\omega,t',x') +
   u^{R *}_H(\omega,t,x) \, u^{L *}_H(\omega,t',x') \right. \nonumber \\
   & & \;\; \left.  + \cosh \left(\frac{\pi \omega}{\kappa}\right) \left[ u^{L}_H(\omega,t,x) \, u^{L *}_H(\omega,t',x')
   + u^{L *}_H(\omega,t,x) \, u^{L}_H(\omega,t',x') \right. \right. \nonumber \\
  & & \;\;\; \left. \left. + u^{R}_H(\omega,t,x) \, u^{R *}_H(\omega,t',x') + u^{R *}_H(\omega,t,x) \, u^{R }_H(\omega,t',x') \right]
   \right\} \label{Ieq}
\eea

\section{Mode Functions}

It is next necessary to obtain explicit expressions for the mode functions.  Their normalization has been described above.  For each of them the mode equation can be solved using separation of variables.  The result is
\bes
\bea  u^{L}_H(\omega,t,x) &=& \frac{1}{\sqrt{4 \pi \omega}} e^{i \omega t} \chi^{L}_H (x) \, \label{uHin}  \\
      u^{R}_H(\omega,t,x) &=& \frac{1}{\sqrt{4 \pi \omega}} e^{-i \omega t} \chi^{R}_H (x) \, \label{uHout} \\
      u^{R}_I(\omega,t,x) &=& \frac{1}{\sqrt{4 \pi \omega}} e^{-i \omega t} \chi^{R}_I (x) \, \label{uIout}\ .
\eea
\label{udefs}
\ees
In all cases the part of the mode function that depends on the coordinate $x$ obeys the equation
\be     \frac{d^2 \chi}{d x^{* \,2}} + \omega^2 \chi +  V_{\rm eff} \chi  = 0 \;. \label{chi-eq} \ee
This equation must be solved numerically in general.  If $c \rightarrow {\rm constant}$ as $x \rightarrow \pm \infty$, then $V_{\rm eff} \rightarrow 0$ in
these limits.  Further, because $c \rightarrow v_0$ at the horizon ($x=0$), $V_{\rm eff} \rightarrow 0$ at the horizon as well.  This aids in determining the
boundary conditions on $\chi$.

The easiest case to consider is $\chi^{L}_H$.  In the {\it in} region, $x^*$ is the time coordinate.  Thus on both the past and future horizons (which are both
at $x = 0$) one has $\chi = e^{- i \omega x^*}$.  Since $\chi^{L}_H$ enters the $L$ region from the past horizon $H^-$ the modes are initially right moving.
This is the reason for the factor of $e^{i \omega t}$ in Eq.~\eqref{uHin}.  For numerical purposes it is useful to break $\chi^{L}_H$ into real and imaginary parts.
Thus we define $ \chi^{L}_c $ to be the solution to Eq.~\eqref{chi-eq} which in the limit $x \rightarrow 0^-$ has the behavior
\bes
\be \chi^{L}_c \rightarrow \cos(\omega x^*) \;. \label{chi-in-c} \ee
Similarly in this limit we define
\be \chi^{L}_s \rightarrow \sin(\omega x^*) \;. \label{chi-in-s} \ee
Then
\be \chi^{L}_H =  \chi^{\rm in}_c + i \chi^{\rm in}_s  \;. \label{chi-in-H}  \ee  \ees
The numerical computation of $\chi^{L}_c$ and $\chi^{L}_s $ is described in more detail in the next section.

In the $R$ region the situation is more complicated.  It is useful to define two solutions to Eq.~\eqref{chi-eq} by their behavior in the large $x$ limit:
\bes \bea  \chi^{R}_c &\rightarrow& \cos(\omega x^*) \;, \\
           \chi^{R}_s &\rightarrow& \sin(\omega x^*) \;.
\eea \ees
Near $x=0$ these solutions will have the form
\bes \bea   \chi^{R}_c &\rightarrow& A \cos(\omega x^*) + B \sin(\omega x^*) \;,  \\
     \chi^{R}_s &\rightarrow& C \cos(\omega x^*) + D \sin(\omega x^*) \;.
\eea \label{ABCD} \ees
Using these, one can define solutions which correspond to right moving and left moving waves in the large $x$ limit as
\bes \bea  \chi^{\infty}_r &=& \chi^{R}_c + i \chi^{R}_s  \;, \label{chi-R}  \\
           \chi^{\infty}_\ell &=& \chi^{R}_c - i \chi^{R}_s  \;. \label{chi-L}
\eea \label{chi-RL} \ees
Near $x = 0$ these solutions will have both right and left moving parts so that
\bes  \bea  \chi^{\infty}_r &\rightarrow& E_r e^{i \omega x^*}  + F_r e^{-i \omega x^*} \;, \label{chi-R-x-0} \\
         \chi^{\infty}_\ell &\rightarrow& E_\ell e^{i \omega x^*}  + F_\ell e^{-i \omega x^*} \;. \label{chi-L-x-0}
\eea \label{chi-RL-x-0} \ees
One easily finds that
\bes \bea  E_r &=&  \frac{1}{2} [ A+D - i(B - C)] \;,  \\
           F_r &=&  \frac{1}{2} [ A-D + i(B + C)] \;,  \\
           E_\ell &=&  \frac{1}{2} [ A-D - i(B + C)] \;,  \\
           F_\ell &=&  \frac{1}{2} [ A+D + i(B - C)] \;.
\eea \label{EFABCD} \ees

For the modes which enter the $R$ region from the past horizon,
the normalization occurs at that horizon while the boundary condition on $\chi^{R}_H$ is that the mode function should be a right moving wave in the limit $x \rightarrow \infty$.
Thus
\be \chi^{R}_H =  N \chi^{\infty}_r  \;, \label{b-1} \ee
The normalization constant $N$ is determined through the normalization condition~\eqref{uH-out-norm} on $H^-$.  Near $x=0$ it is
the right moving part of $\chi^{R}_H$ which corresponds to the part of the mode function coming from $H^-$. Using Eqs.~\eqref{uHout} and~\eqref{chi-R-x-0} one finds that
\be N = \frac{1}{E_r}  \label{Ndef} \;. \ee

To find the behavior of $\chi^{R}_H$ for $x <0$ we note that it is the left moving part near $x = 0^+$ which goes through the future horizon. Then by continuity of
the total mode function $u^{R}_H$ across the future horizon one finds that for $x < 0$
\be \chi^{R}_H = \frac{F_r}{E_r} \left( \chi^{L}_c - i \chi^{L}_s \right) \;. \label{chi-out-H-x-lt-0} \ee

Finally we discuss the modes $u^{R}_I$ which originate on $I^-$ in the $R$ region.  The normalization for these modes occurs on $I^-$ and is given in Eq.~\eqref{uI-norm}.
Because of backscattering part of the mode function reaches $I^+$.  Therefore,
\be \chi^{R}_I = \chi^{ \infty}_\ell + K \chi^{\infty}_r \label{chi-I} \;, \ee
with $K$ a constant.
The boundary condition is that on $H^-$, $u^{R}_I = 0$.  This is accomplished by having $\chi^{R}_H$ be a left moving wave in the limit $x \rightarrow 0^+$, that is
\be  \chi^{R}_I = G e^{-i \omega x^*}  \;, \label{chi-I-x-0}  \ee
with $G$ another constant.  Using Eqs.~\eqref{chi-RL-x-0} in Eq.~\eqref{chi-I} and setting the result equal to~\eqref{chi-I-x-0} gives
\bes \bea  K &=& - \frac{E_\ell}{E_r}  \\
           G &=& F_L - \frac{E_\ell F_r}{E_r} \;.
\eea \ees
For $x < 0$ we again use continuity of the mode function at $H^+$ to obtain
\be \chi^{R}_I = \left(F_\ell - \frac{E_\ell F_r}{E_r} \right) \left( \chi^{L}_c - i \chi^{L}_s \right) \;. \label{chi-out-I-x-lt-0} \ee

This ends the derivation of the equations necessary for the computer program.  We used Mathematica to combine them to give expressions for the density-density correlation function
in terms of the modes $ \chi^{L}_c$,  $ \chi^{L}_s$, $ \chi^{R}_c$, $ \chi^{R}_s$ and their derivatives and also in terms of the constants $A$, $B$, $C$, $D$.
All of these quantities are numerically computed.

\section{Numerical Computations}
\label{numerical}

In this section we discuss the numerical computation on the density density correlation function.  This involves numerical computations of the mode functions in both the $L$ and $R$ regions.  The results are substituted into the expressions for the density density correlation function which involve integrals over the frequency $\omega$.  Because of the different behaviors of the mode functions in the {\it in} and {\it out} regions there are different specific expressions for the density density correlation function in terms of the mode functions and their derivatives depending on where the two points are located.  These have been obtained using the algebraic manipulation program Mathematica but they are too long to be shown here.

There is a problem that arises in computing the mode integrals in all but one case.  That is that the process of computing the integrals over $\omega$ and computing the derivatives of the two point function necessary to obtain the density density correlation function do not commute.  The point is that for the two point function the integrals over $\omega$ are well defined and converge in the limit $\omega \rightarrow \infty$ so long as the points are separated.  However, if one takes the relevant derivatives first and thus obtains an integral over derivatives of the mode functions, then the resulting integrals over $\omega$ are not well defined.  This is discussed in more detail in Sec.~\ref{modes-integrals} where it is shown that a subtraction procedure can be used to overcome this problem.  Note that there is also the usual infrared divergence of the two point function in two dimensions, but this can be taken care of with a judicious choice of infrared cutoff, which in itself is reasonable because of the finite dimensions of the physical system being modeled.

\subsection{Numerical Calculations of the Radial Mode Functions}
\label{modes-numerical}

In both the $L$ and $R$ regions the numerical computation of the radial mode functions can be accomplished by computing the functions $\chi_c$ and $\chi_s$ for these regions.
In the $R$ region these mode functions asymptotically approach $\cos \omega x^*$ and $\sin \omega x^*$ as $x \rightarrow \infty$.  In the $L$ region they have the same asymptotic behaviors in the limit $x \rightarrow 0^-$.  The starting values for these modes can be made arbitrarily accurate by starting at a large enough value of $x$ for the $R$ modes and a value of $x$ close enough to zero for the $L$ modes.

The only other task that needs to be accomplished before the mode functions are substituted into the expressions for the density density correlation function is that the parameters $A$, $B$, $C$, and $D$ must be numerically determined.  This is accomplished by matching the numerically computed mode functions $\chi^{R}_c$ and $\chi^{R}_s$ to Eqs.~\eqref{ABCD}.
One way to do this is to do the matching at a number of values of $x$ with $0 < x \ll 1$ and then use a linear extrapolation routine to find the values that $A$, $B$, $C$, and $D$ in the limit that $x \rightarrow 0$.

\subsection{Numerical Computation of the Mode Integrals}
\label{modes-integrals}

The computation of the mode integrals in the density density correlation function would be a straight-forward numerical exercise if they converged.
However there are two problems.  One is that because we are working with an effective relativistic two dimensional quantum field theory there is an
infrared divergence in the two point function and therefore there are infrared divergences in the density density correlation function.  The other
is related to the fact mentioned previously that strictly speaking one must compute the mode integral in the two point function and then take
the required derivatives of it to obtain the density density correlation function.

\subsection*{Infrared divergences}

The simplest way to numerically remove the infrared divergence is to impose a lower limit cutoff $\lambda_\ell$ on the mode integrals.   From Eqs.~\eqref{IplusJ},~\eqref{Jdef},~\eqref{Ieq},~\eqref{udefs}, and~\eqref{chi-eq} one can see that the two point function has infrared divergences which go like both $1/\lambda_\ell$ and $\log(\lambda_\ell)$ for a small enough infrared cutoff $\lambda_\ell$.
As shown in Eq.~\eqref{G2formula} the density density correlation function is composed of terms containing two derivatives of the two point function.  Each time derivative brings down a factor of $\omega$ while each space derivative results in one term in which a factor of $\omega$ is brought down, one term in which one or more derivatives of the sound speed $c(x)$ are found, and one term in which a derivative of the spatial part of one of the mode functions occurs.  If the sound speed at the point where the spatial derivative is taken is constant then in all surviving terms a factor of $\omega$ is brought down.  If the sound speed is not constant then there are one or more terms in which either no factor of $\omega$ or only one factor of $\omega$ are brought down by the spatial derivatives.  Thus if both points of the density density correlation function are in regions where the sound speed is constant (and therefore the effective potential is zero) then there are no infrared divergences in the density density correlation function.  Otherwise infrared divergences of the types discussed above occur.

Even if there are no infrared divergences, the finite physical size of the system means that there will be an infrared cutoff.  As discussed above for low enough values of the cutoff there will be terms containing the cutoff which go like $1/\lambda_\ell$ and $\log(\lambda_\ell)$.  However if a higher value is given to the cutoff then oscillatory behavior in the density density correlation function is generated by the cutoff.  To see why this occurs one can look at the behaviors of the mode functions when the effective potential is zero.  Then it is not hard to show that there will be terms which go like ${\rm ci}(\lambda_\ell \Delta u)$,  ${\rm ci}(\lambda_\ell \Delta v)$, and so forth.  If one point is outside the horizon near one end of the BEC and the other is inside the horizon near the other end of the BEC then roughly speaking $\Delta u \sim \Delta v \sim \Delta x$, with $\Delta x$ the size of the BEC.  Since the sound speed is of order unity in the models we consider it is clear that in this case the arguments of the cosine integral functions are of order unity.  This leads to oscillatory behavior in terms of the dependence on the value of the cutoff $\lambda_\ell$.  However, the approximations we are using assume a BEC of infinite length and so are not expected to be valid near the edges of the BEC where other effects that are being neglected may be important.  Therefore we shall restrict consideration to much smaller values of $\Delta x$ such that $\lambda_\ell |\Delta x| \ll 1$.
The effects of the infrared cutoff on the density density correlation function when this restriction is satisfied are discussed further in Sec.~\ref{results}.

\section*{Large frequency behavior}

When computing the density density correlation function numerically one must first find numerical values for the mode functions and their derivatives and
then substitute these into the appropriate expression for the correlation function and numerically compute the integrals.  However, if this is done
one finds that some of the resulting integrals do not converge.  Instead for large values of the frequency $\omega$ their integrands oscillate in $\omega$
but with either growing or approximately constant amplitudes.  An illustration of what is happening can be obtained from the cosine integral function.
\be  {\rm ci}(\lambda x) = - \int_\lambda^\infty d \omega \frac{\cos x \omega}{\omega} \;. \ee
If one computes the derivative with respect to $x$ of both sides and interchanges the order of integration and differentiation on the right hand side then
\be \frac{\cos (\lambda x)}{x}  =  \int_\lambda^\infty d \omega \sin(x \omega) \;. \ee
The right hand side of this equation is clearly not well defined since the amplitude of the integrand is constant in the limit $\omega \rightarrow \infty$.

One way around this problem is to evaluate the integrands for the integrals in the density density correlation function in the large $\omega$ limit.  Then
the terms that are poorly behaved at large $\omega$ can be identified and subtracted off.  Because it is the large $\omega$ limit, this can be done analytically and then
the specific terms that are subtracted off can be added back and treated analytically.  A simple example of how this works would be
\be \int_\lambda^\infty d \omega \frac{\cos(\omega x)}{\sqrt{1 + \omega^2}} = \int_\lambda^\infty d \omega \left( \frac{\cos(\omega x)}{\sqrt{1 + \omega^2}} - \frac{\cos(\omega x)}{\omega} \right) \, - {\rm ci}(\lambda x) \;. \ee
Then if one wanted the first derivative of this with respect to $x$ one would have
\be \frac{d}{d x} \int_\lambda^\infty d \omega \frac{\cos(\omega x)}{\sqrt{1 + \omega^2}} = -\int_\lambda^\infty d \omega \left(\omega \frac{\sin(\omega x)}{\sqrt{1 + \omega^2}} - \sin(\omega x) \right) + \frac{\cos(\lambda x)}{x}  \;. \ee
In this case the integral on the right hand side has an integrand that oscillates but with an amplitude that vanishes in the limit $\omega \rightarrow \infty$.  It can therefore be computed numerically.

To evaluate the large $\omega$ behaviors of the integrands of the integrals that occur in the density density correlation function one can examine the behaviors of solutions
to the mode equation in this limit.  It turns out that these can be obtained by using a Volterra series to find solutions for the mode functions in the $L$ and $R$ regions.
To solve Eq.~\eqref{chi-eq} in terms of a Volterra series one first notes that $V_{\rm eff}$ vanishes in the limit $x \rightarrow 0$.  If $\chi^{L}_0$ is any solution to the equation when $V_{\rm eff} = 0$ then a formal solution to the full equation is
\be \chi^{L}(x) = \chi^{L}_0(x) - \frac{1}{\omega} \int_{-\infty}^{x^*} d y^* \sin[\omega(x^*-y^*)] V_{\rm eff}(y^*) \chi^{L}(y^*) \;.  \label{chi-formal-in} \ee
 This equation can be solved by iteration.  After the first iteration one finds
\be \chi^{L}(x) = \chi^{L}_0(x) - \frac{1}{\omega} \int_{-\infty}^{x^*} d y^* \sin[\omega(x^*-y^*)]  V_{\rm eff}(y^*) \chi^{L}_0(y^*) \;.  \label{chi-first-in} \ee
In the {\it out} region $V_{\rm eff}$ vanishes in the limit $x \rightarrow \infty$.  In this case if $\chi^{R}_0$ is a solution when $V_{\rm eff} = 0$ then the formal solution is
\be \chi^{R}(x) = \chi^{R}_0(x) + \frac{1}{\omega} \int_{x^*}^\infty d y^* \sin[\omega(x^*-y^*)]  V_{\rm eff}(y^*) \chi^{R}_(y^*) \;,  \label{chi-formal-out} \ee
After one iteration
\be \chi^{R}(x) = \chi^{R}_0(x) + \frac{1}{\omega} \int_{x^*}^\infty d y^* \sin[\omega(x^*-y^*)]  V_{\rm eff}(y^*) \chi^{R}_0(y^*) \;.  \label{chi-first-out} \ee

To obtain the large $\omega$ limit of the mode functions we note that if $V_{eff} = 0$ then
\bes \bea \chi^{L}_c &=& \cos(\omega x^*)  \;, \label{chicV0in} \\
     \chi^{L}_s &=& \sin(\omega x^*)  \;, \label{chisV0in} \\
\chi^{R}_c &=& \cos(\omega x^*)  \;, \label{chicV0out} \\
     \chi^{R}_s &=& \sin(\omega x^*)  \;. \label{chisV0out}
\eea  \ees
Substituting these expressions into Eqs.~\eqref{chi-first-in} and~\eqref{chi-first-out}, and using simple trigonometric identities one finds that to $O(1/\omega)$
\bes \bea \chi^{L}_c &=& \cos(\omega x^*) - \frac{\sin(\omega x^*)}{2 \omega} \int_0^x dy V(y)  \;,  \label{chicVin}  \\
           \chi^{L}_s &=& \sin(\omega x^*) + \frac{\cos(\omega x^*)}{2 \omega} \int_0^x dy V(y)  \;,  \label{chisVin}  \\
 \chi^{R}_c &=& \cos(\omega x^*) + \frac{\sin(\omega x^*)}{2 \omega} \int_x^\infty dy V(y)  \;,  \label{chicVout}  \\
\chi^{R}_s &=& \sin(\omega x^*) - \frac{\cos(\omega x^*)}{2 \omega} \int_x^\infty dy V(y)  \;.  \label{chisVout}
\eea \ees

One can use these results to obtain the large $\omega$ limits of the matching parameters in Eqs.~\eqref{ABCD}.  However from Eqs.~\eqref{EFABCD} it can be seen that what one really needs are the sums and differences of $A$ and $D$, and of $B$ and $C$. To $O(\omega^0)$ one finds that $A = D = 1$ and $B = C = 0$.  To $O(\omega^{-1})$ the sums and differences are
\bea A + D & = & 2 \;, \nonumber \\
     A - D &=& 0 \;, \nonumber \\
     B + C &=& 0 \;, \nonumber \\
     B - C &=& \frac{1}{\omega} \int_0^\infty dy V(y) \;.
\label{ABCDdif}
\eea

These results can be substituted into the integrals for the density density correlation function in terms of the modes $\chi$ and the matching parameters $A$, $B$, $C$, and $D$ to
obtain an approximate expression which is valid in the large $\omega$ limit.  This is then subtracted from the density density correlation function and added back on.  The part
that is added back on is computed analytically by first computing the approximate two point function and then taking the relevant derivatives.  In the rest one has integrals which
can now be computed numerically because they are well behaved in the limit $\omega \rightarrow \infty$.

\section{Numerical Results}
\label{results}

Numerical computations of the density-density correlation function have been carried out for the sound speed profile
\be c = \sqrt{c_1^2+ \frac{1}{2}(c_2^2-c_1^2)\left[1 + \frac{2}{\pi} \tan^{-1}\left( \frac{x+b}{\sigma_v} \right) \right]}
 \label{c-profile} \ee
with
\be b = \sigma_v \tan\left[\frac{\pi}{c_2^2-c_1^2} \left(v_0^2-\frac{1}{2} (c_1^2+c_2^2)\right)\right] \;.
\label{horizon-shift} \ee
  The values used for some of the constants were
\bea
     v_0 &=& \frac{3}{4} \;, \nonumber \\
     c_1 &=&  \frac{1}{2} \;, \nonumber \\
     c_2 &=& 1 \;. \nonumber \\
\eea
This is the same type of profile as that used in Ref.~\cite{paper2} where numerical calculations of the density density correlation function were carried out in the context of condensed matter physics.  However for those calculations $c_1 = 1$ and $c_2 = 1/2$ so that the BEC was moving to the right, not the left~\cite{iacopo-private}.  Also $b = 0$ so that the horizon was not at $x = 0$.

Calculations were done for
 several values of the parameter $\sigma_v$ ranging from $1/4$ to $8$.  Typical values used for the infrared cutoff $\lambda$ in the integrals over the frequency $\omega$ ranged from $2 \times 10^{-6}$ to $2 \times 10^{-3}$.  All of the plots shown here have an infrared cutoff of $\omega = 2 \times 10^{-6}$.

A careful analysis of~\eqref{c-profile} shows that the derivatives of $c(x)$ are larger near the horizon for smaller values of $\sigma_v$.  The hydrodynamic approximation that we are using is valid only if the derivatives of $c(x)$ are relatively small.  For the profile in~\eqref{c-profile} the hydrodynamical approximation was shown in~\cite{paper2} to be valid when $\sigma_v \stackrel{_>}{_\sim} 4$.  Thus while we have computed the density density correlation function for smaller values of $\sigma_v$ we display our results only for $\sigma_v = 8$.

In all cases the correlation function was computed for equal values of the lab time $T$ and in fact specifically for $T=0$.  The arbitrary constants in Eq.~\eqref{tTout} and in Eq.~\eqref{xstarout} for the {\it out} region have been given the values $x_1 = x_3 = 1$.  In the {\it in} region the corresponding constants were given the values $x_2 = x_4 = -1$.  To have continuity of the coordinate $v = t + x^*$ across the future horizon it is easy to show by equating the expressions for $v$ in the {\it in} and {\it out} regions at $x = 0$ that
\be a = -\int_{-x_0}^{x_0} \frac{dy }{c(y)+ v_0}  \;.  \label{a-def} \ee

As can be seen in Eqs.~\eqref{two-point-1} and~\eqref{Ieq}, the two point function can be written in terms of an integral ($I$) over the mode functions $u^{\rm in}_H$ and $u^{\rm out}_H$ and an integral ($J$) over the mode functions $u^{\rm out}_I$.  Thus the density density correlation function can also be separated into terms containing integrals over $u^{\rm in}_H$ and $u^{\rm out}_H$ and terms containing integrals over $u^{\rm out}_I$.

In Ref.~\cite{paper1} the density density correlation function was computed analytically with the following assumptions and restrictions:  It was computed with one point inside the horizon and one point outside.  The assumption was made that the point inside was in a region where the sound speed was constant and the point outside was in a region where the sound speed was also constant but had a different value.
Having neglected the effective potential $V_{eff}$ in Eq.~(\ref{wave-eq-2}) and hence the backscattering, the computed correlation corresponds to that in Fig.~\ref{fig9} (between the modes in Fig. ~\ref{fig6} and Fig. ~\ref{fig7}).
The main result of that paper is the existence of a negative correlation peak (i.e. a trough) in the correlation function which would not be there if there was no horizon.

In Ref.~\cite{paper2} a fully quantum mechanical computation of the density  density correlation function was done and a comparison was made with the result of~\cite{paper1}.  The existence of the negative correlation peak when one point is inside and one point is outside the horizon was confirmed.  The peak value of the correlation function along with the full width at half maximum of the peak were compared when the points were far enough from the horizon that the sound speed was effectively constant.  They were found to be in approximate agreement only for $\sigma_v \stackrel{_>}{_\sim} 4$.  Also found were two other correlation peaks, both of which are substantially weaker than the one found in~\cite{paper1}. One of these correlation peaks is negative (between the modes in Fig. ~\ref{fig7} and Fig. ~\ref{fig8} modes, see Fig.~\ref{fig11}).  It occurs when both points are inside the horizon.  The other is a very weak (positive) correlation peak which occurs when one point is outside and one point is inside the horizon (between the modes in Fig. ~\ref{fig6} and Fig. ~\ref{fig8}, see Fig.~\ref{fig10}).

\begin{figure}
\vskip -0.2in \hskip -0.4in
\includegraphics[scale=0.3,angle=0,width=2.4in,clip]{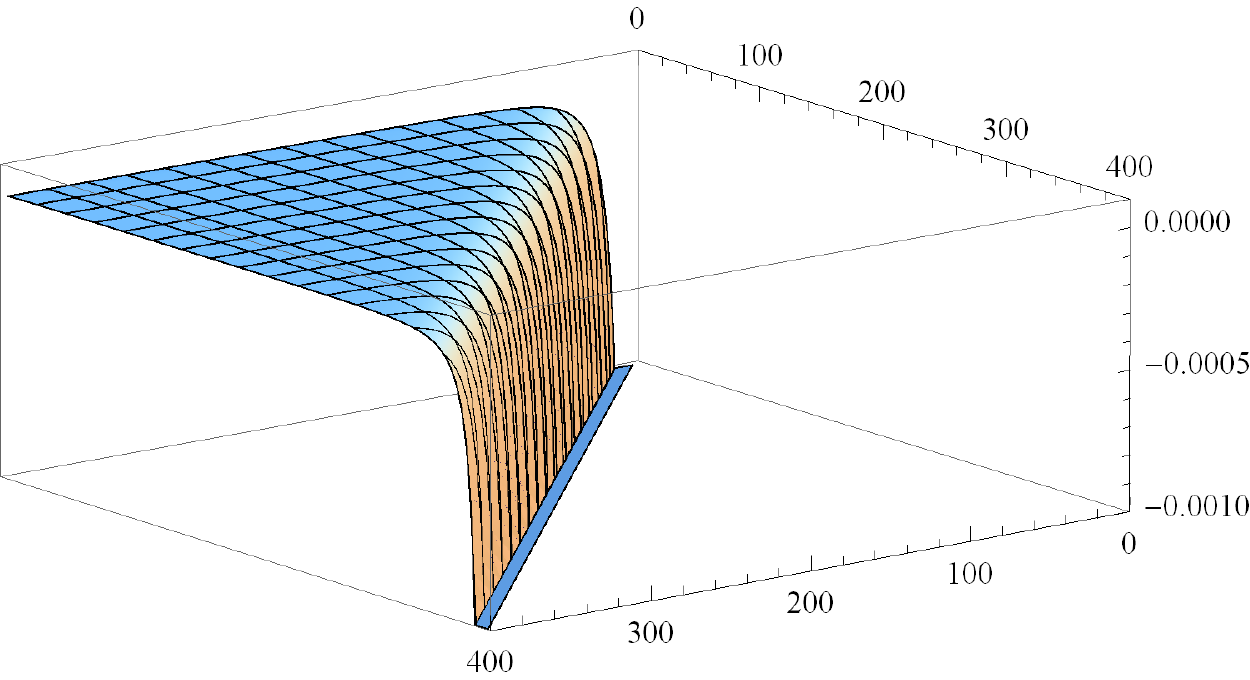}
\includegraphics[scale=0.3,angle=0,width=2.4in,clip]{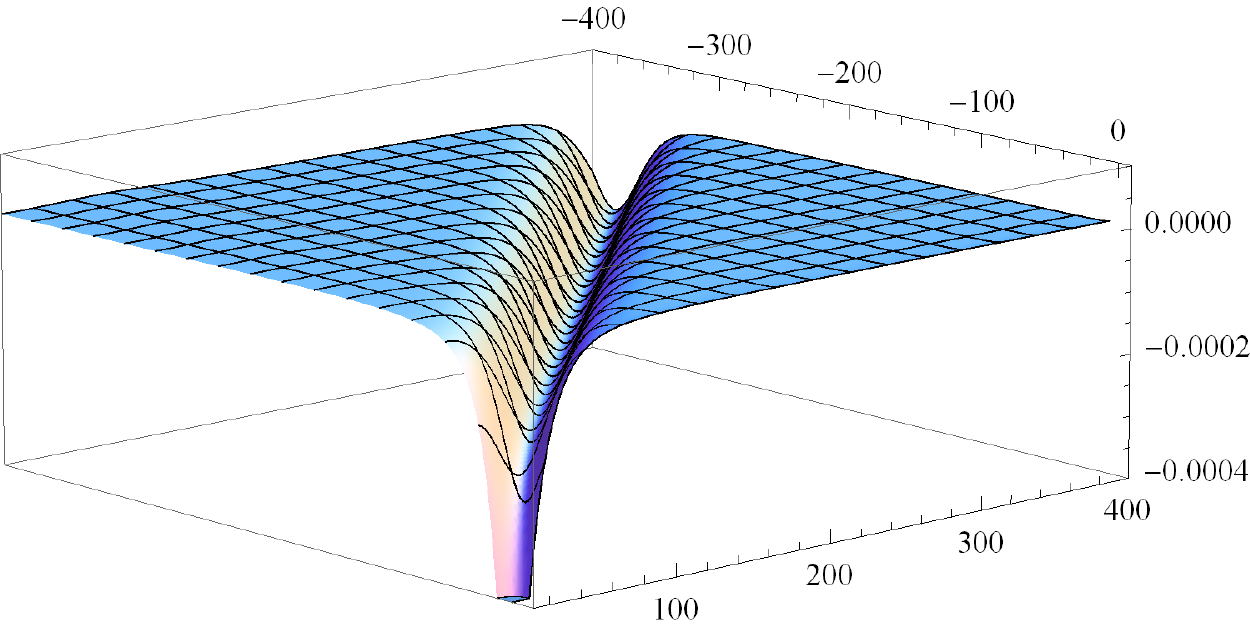}
\begin{center}
\includegraphics[scale=0.3,angle=0,width=2.4in,clip]{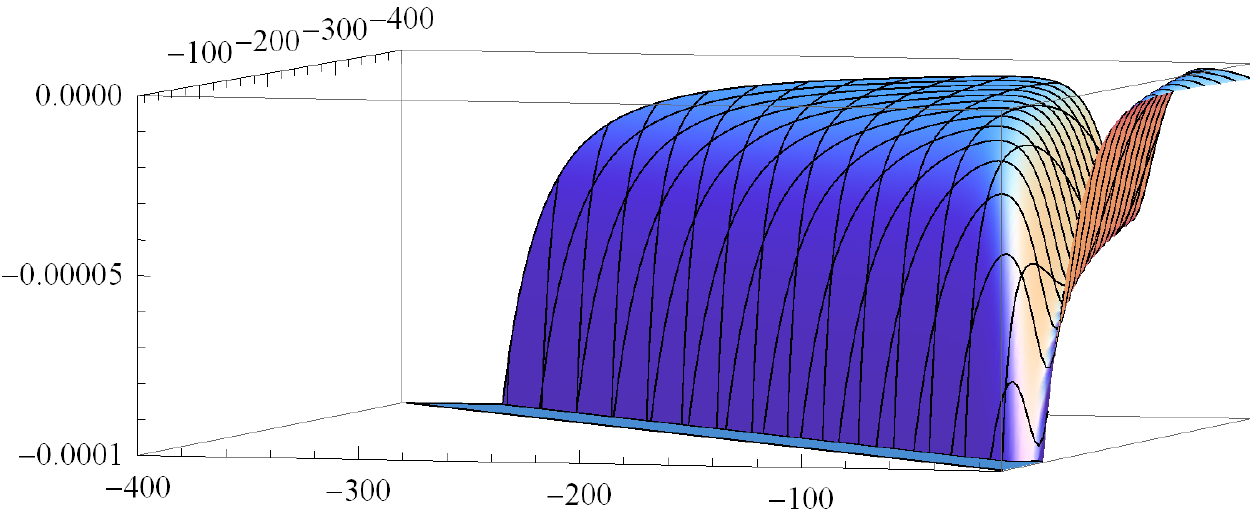}
\end{center}
\vskip -.2in \caption{The density density correlation function is shown for $\sigma_v = 8$. Both points are outside the horizon in the top left plot, one point is inside and one point is outside the horizon in the top right plot, and both points are inside the horizon in the bottom plot.  In all cases the lab time $T$ is the same for both points and and an infrared cutoff in the frequency integrals of $\lambda = 2 \times 10^{-4}$ has been used.  Note that in the top left and bottom plots the bottom looks flat.  This is a distortion caused by imposing a limit on the range of values of the correlation function that are plotted. }
\label{main-plots}
\end{figure}

Our results for the density density correlation function when $\sigma_v = 8$ are shown in Fig.~\ref{main-plots}.  When both points are inside or both points are outside the horizon there is a divergence in the correlation function when the points come together.  Also the correlation function is symmetric about the point separation.  Thus only half the range of values of the points is shown in those plots.  There is no divergence in the density density correlation function and it is not symmetric about the point separation when one point is inside and the other point is outside the horizon.  Thus we show the full range of values of the points in this case.

When both points are outside the horizon the only feature we observed in the density density correlation function is the divergence as the two points come together.

\begin{figure}
\vskip -0.2in \hskip -0.4in
\includegraphics[scale=0.3,angle=0,width=2.4in,clip]{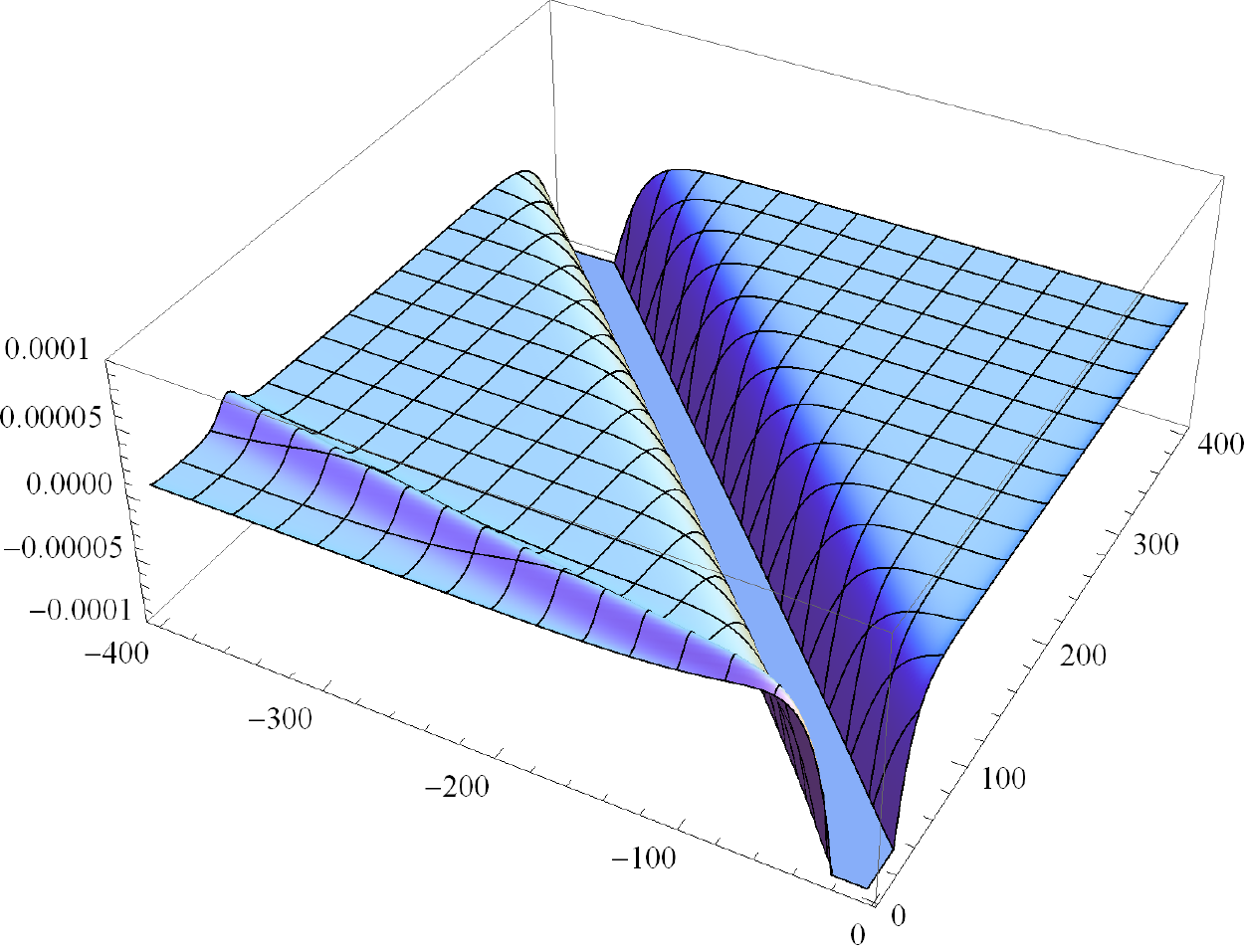}
\includegraphics[scale=0.3,angle=0,width=2.4in,clip]{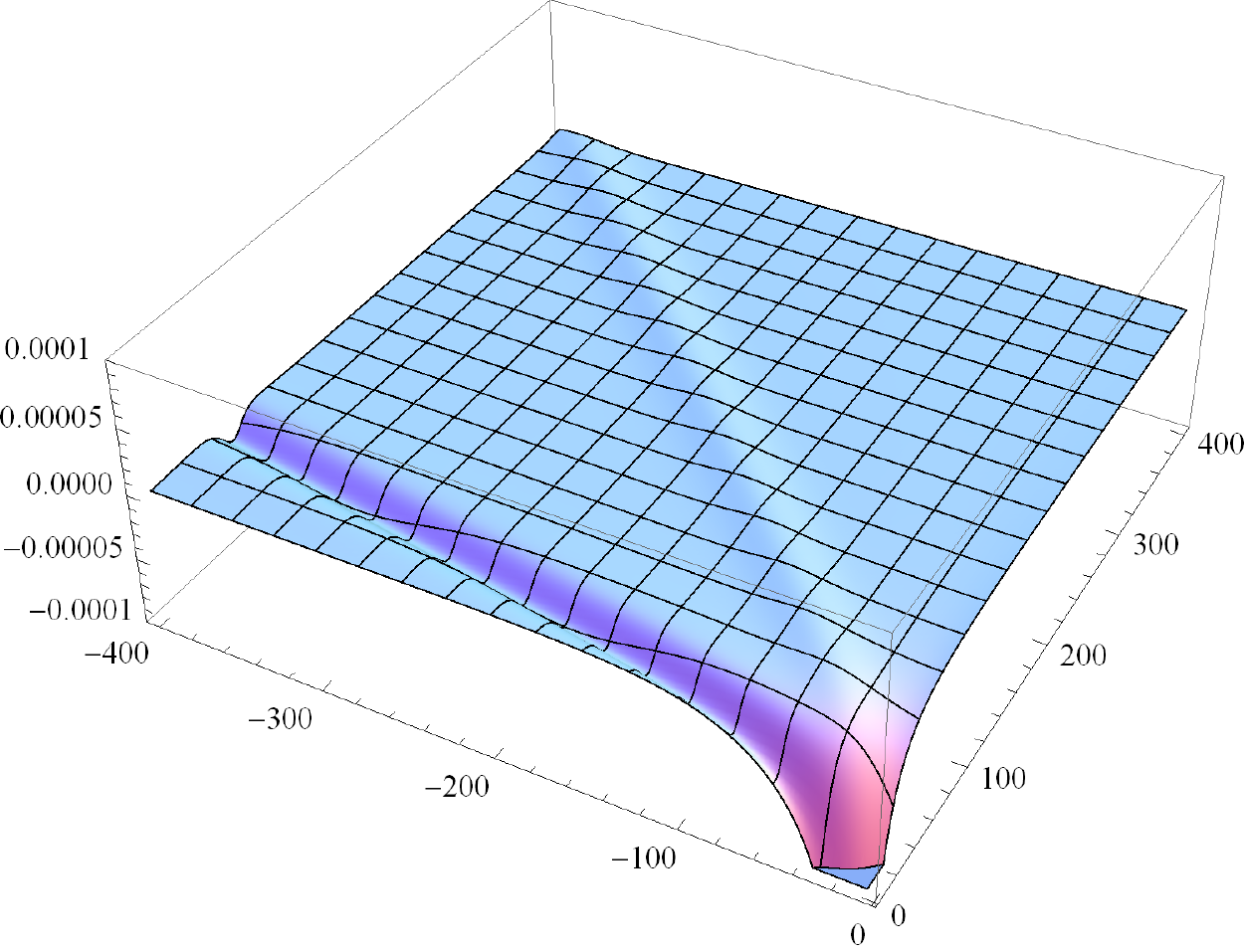}
\begin{center}
\includegraphics[scale=0.3,angle=0,width=2.4in,clip]{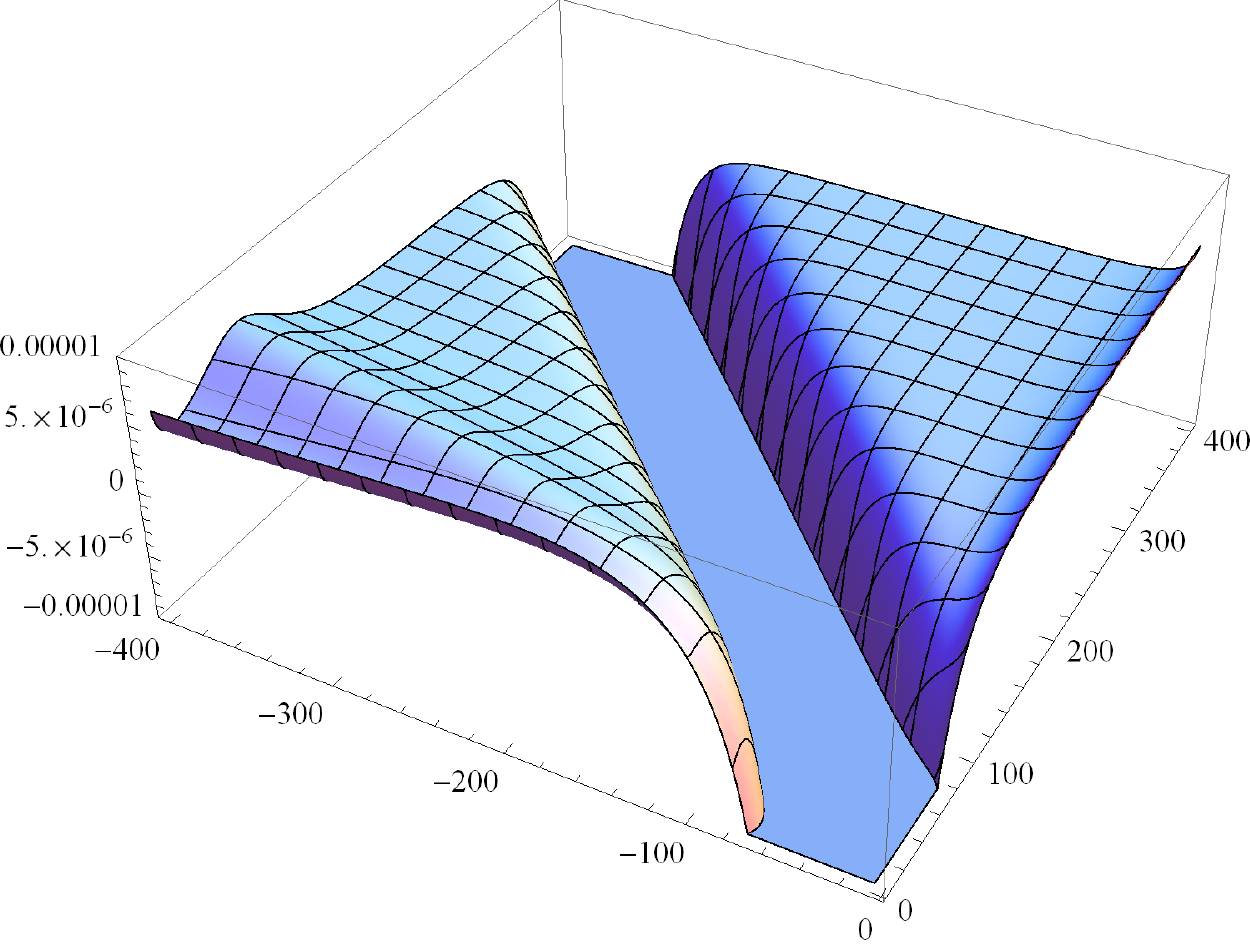}
\end{center}
\vskip -.2in \caption{The density density correlation function is shown for $\sigma_v = 8$ when one point is inside and one point is outside the horizon.  The plot on the top left shows the contribution from the $u^{\rm in}_H$ and $u^{\rm out}_H$ modes.  The plot on the top right shows the contribution from the $u^{\rm out}_I$ modes.  The plot on the bottom shows the total.
In all cases the lab time $T$ is the same for both points and and an infrared cutoff in the frequency integrals of $\lambda = 2 \times 10^{-4}$ has been used.  The vertical scale is set so that the positive correlation peak on the upper left plot, the two maxima on the upper right plot, and the positive correlation peak in the bottom plot can be seen.  This has the effect of distorting the shape of the negative correlation peak in the plots on the top left and bottom so that it appears flat near its maximum.  Note that the vertical scale of the bottom plot is $1/10$ as large as the top plots.  It is necessary to do this to clearly show the small peak because of the partial cancelation in this peak that occurs when the two contributions are added together. }
\label{H-I-in-out}
\end{figure}

When one point is inside and one point is outside the horizon we find the same negative correlation peak as was found when the potential is zero.  It it interesting in this case to look separately at the contribution of the $u_H^L,\ u_H^R$  modes and those of the $u^{R}_I$ modes.  These are shown in the first two plots in Fig.~\ref{H-I-in-out}.   Note that the contribution from the $u^{R}_I$ modes is actually a positive correlation peak.  However, it is significantly smaller in magnitude than the negative correlation peak from the other set of modes.  The effect on the magnitude of the negative correlation peak by including the effective potential in the mode equations and including the contribution from the $u^{R}_I$ modes was observed to be small, usually less than $10\%$.

As can be seen in the bottom plot of Fig.~\ref{H-I-in-out} there is also a small positive correlation peak which occurs when one point is inside and the other point is outside the horizon.  This one lies closer to the horizon than the larger negative one and its maximum value is significantly smaller than that of the main peak.
This appears to correspond to the correlator between the modes in Fig. ~\ref{fig6} and Fig. ~\ref{fig8}, see Fig.~\ref{fig10}.
A careful examination of Fig.~\ref{H-I-in-out} shows that the contribution from the $u^{\rm in}_H$ and $u^{\rm out}_H$ modes is a positive correlation peak and the contribution from the $u^{\rm out}_I$ modes is a negative correlation peak.  However, the two only partially cancel when added together and the result is the small positive correlation peak in the bottom plot.

\begin{figure}
\vskip -0.2in \hskip -0.4in
\includegraphics[scale=0.3,angle=0,width=2.4in,clip]{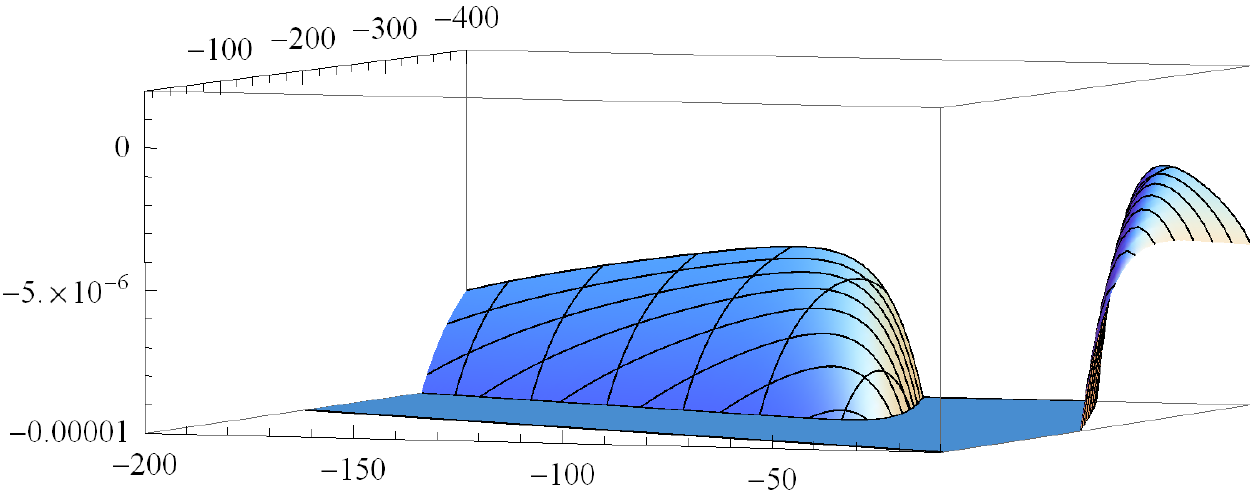}
\vskip -.2in \caption{The density density correlation function is shown for $\sigma_v = 8$ when both points are inside the horizon.  This is the same as the bottom plot in Fig.~\ref{main-plots} but the vertical scale has been changed to more clearly show the maxima on either side of the negative correlation peak.  This has the effect of distorting the shape of the larger negative correlation peak so that it appears flat near its maximum.}
\label{in-in}
\end{figure}

In the bottom plot of Fig.~\ref{main-plots} one finds a negative
correlation peak which occurs when both points are inside the horizon.
This corresponds to the correlator between the modes in Fig.~\ref{fig7} and
Fig.~\ref{fig8}, see Fig.~\ref{fig11}.
If the vertical scale is decreased by a factor of ten in order to show more
detail then as shown in Fig.~\ref{in-in} the density density correlation
function reaches a maximum value on either side of the negative
correlation peak.  The outer maximum appears to be due to the existence of
the negative correlation peak just mentioned and also the very large one
that occurs when the points come together.  There must be a maximum in
between.  Something similar may happen for the other maximum which is in
between the negative correlation peak and the horizon.  However, we are not
able to numerically compute the density density correlation function when
one point is on the horizon so we don't know its value there.

When one point is inside and one point is outside the horizon, a quantitative comparison of our numerical results with the analytic calculation in~\cite{paper1} gives agreement to within approximately $10 \%$ or better for the size of the negative correlation peak for all values of $\sigma_v$ that were considered.  As mentioned previously the comparison of the numerical computations done in~\cite{paper2} with the analytic calculation in~\cite{paper1} showed approximate agreement for $\sigma_v \stackrel{_>}{_\sim} 3$.  For smaller values the size of the peak predicted by the analytic calculation becomes much larger than that found in the numerical calculations in~\cite{paper2}.  Since our numerical results take the effective potential $V_{\rm eff}$ in the mode equation into account, they give the correct quantum field theory in curved space prediction for the density density correlation function.  Our numerical calculations thus confirm  that the discrepancy is due to the fact that the hydrodynamical approximation breaks down for $\sigma_v \stackrel{_<}{_\sim} 3$ due to the rapid variation of the sound speed near the horizon for those values.

  From the above discussion it is clear that our numerical results reproduce all of the same late time peaks in the correlation function that were found in~\cite{paper2,paper3,rpc}.  This is true even for $\sigma_v \stackrel{_<}{_\sim} 3$.  In~\cite{paper2} only the results for the negative correlation peak when one point is inside and one point is outside the horizon were shown for more than one value of $\sigma_v$.  For the other peaks the results shown in~\cite{paper2,paper3,rpc} were only shown for the case $\sigma_v = 1/2$. However, we have been given access~\cite{iacopo-private} to some of the numerical data that was generated for~\cite{paper2}.  This has allowed quantitative comparisons to be made between our numerical results and those numerical results for the other correlation peaks.

  In the numerical data we were given~\cite{iacopo-private} we could only unambiguously identify the positive correlation peak when one point is inside and one point is outside the horizon for $\sigma_v \le 4$ although there was evidence for it for larger values of $\sigma_v$.  In those cases our value for the size of the peak was always somewhat larger.  However, as discussed below the effects of the infrared cutoff are large enough that the size of this peak can be significantly affected.

  For the negative correlation peak which occurs when both points are inside the horizon we find that at $\sigma_v \ge 4$ there is agreement to roughly $25 \%$  or better between the two data sets with that agreement being better for larger values of $\sigma_v$ as expected.  For all the values of $\sigma_v$ considered our results give a larger value for the size of the negative correlation peak than those for~\cite{paper2}.

  As mentioned above, and as seen in Fig.~\ref{in-in}, when both points are inside the event horizon we find evidence for a maximum in the density density correlation function on either side of the negative correlation peak.  In the numerical data we were given~\cite{iacopo-private} the maximum that is farther from the horizon than the negative peak is always seen.  There is also always evidence for the maximum we see that is closer to the horizon but for $\sigma_v \ge 2$ an actual maximum is not seen.  Instead the value of the correlation function just keeps increasing as one point approaches the horizon (and the other point remains far from the horizon).   For $\sigma_v = 1$ a maximum is seen.  In some cases for our data the inner maximum has a positive value and is thus a positive correlation peak.  The outer maximum however has a negative value except for $\sigma_v =1$ where it has a very small positive value.  In the other data set the outer maximum is less than zero for $\sigma_v \le 6$ and greater than zero for larger values of $\sigma_v$.  For $\sigma_v \ge 1$ the outer maximum is smaller than the inner maximum.  Neither maximum was seen in~\cite{paper2} but both of them can be seen in~\cite{rpc}, although only the larger, inner one is identified as a correlation peak.

As discussed previously the infrared cutoff that we use does have an effect on the value of the density density correlation function and will have a strong effect for values of the cutoff that are either too large or too small.  Different values of the infrared cutoff were considered for some of our calculations.  It was found for example for $\sigma_v = 8$ that when a comparison was made between calculations with a cutoff of  $\lambda = 2 \times 10^{-4}$ and a calculation with a cutoff of $\lambda = 2 \times 10^{-6}$ that the differences were typically in the range $(2 - 4) \times 10^{-6}$.  This is small enough that the negative correlation peaks are not significantly affected.  However, the positive correlation peak and the two maxima discussed above are close enough to zero that their values can be affected.  A similar comparison between the cutoffs of $\lambda = 2 \times 10^{-3}$ and $\lambda = 2 \times 10^{-4}$ yields differences that typically range from $ (1-4) \times 10^{-5}$ which is about an order of magnitude higher.  Thus a cutoff of $\lambda = 2 \times 10^{-3}$ is probably too high.

\section{Conclusions}
\label{Conclusions}

We have developed a method to numerically compute the density density correlation function for a BEC which serves as an analogue black hole. In the process
we have shown that it is necessary to incorporate an infrared cutoff in the frequency.  So long as this cutoff is small in the sense that its product with the
point separation is small, then it does not significantly affect the main peaks or troughs in the correlation function.  However it does significantly affect
the value of the correlation function in regions where the magnitude of that function is small.

In order to correctly compute the correlation function it is necessary to consider modes of arbitrarily high frequencies.  This is exactly what one expects when working with an effective field theory~\cite{donoghue}.  Unless this is done ultraviolet cutoff effects dominate the structure of the correlation function.

To take these modes into account it was found that a type of regularization must be used in order to compute the correlation function numerically.  One must subtract off terms whose amplitude grows or stays the same as the frequency increases.  Then these terms are added back again and the integral over $\omega$ is computed formally by starting with an integral of the form
    \be {\rm ci} \sim \int  d \omega  \cos(\omega u) /\omega \ee
      or
      \be {\rm si} \sim \int  d \omega  \sin(\omega u) /\omega \ee
     and taking derivatives with respect to u.
This reason for this is related to the fact that for the original correlation function one must first compute the integral over the mode frequencies and then take the derivatives
necessary to compute the density density correlation function.  These two operations do not commute.  But if one performs a regularization as just described then there is no
problem in first computing the derivatives and then integrating over the frequencies. This is the desired order of operations for numerical computations.

Our numerical results show that for the models considered the size and location of the main negative peak in the correlation function when one point is inside and one point is outside the horizon is not changed significantly by including the potential.  However, including the potential does correctly reproduce the features which were found in the condensed matter calculations in~\cite{paper2}, namely  a negative correlation peak which occurs when both points are inside the horizon
and a small positive correlation peak that occurs when one point is inside and the other point is outside the horizon.
Finally we find two maxima which occur on either side of the negative correlation peak when both points are inside the horizon.  These maxima do not appear on the plots of Ref.~\cite{paper2}, but they do appear in the plot in~\cite{rpc} although only the one closer to the horizon is identified as a correlation peak.
\par In cases where the hydrodynamical approximation is valid there is reasonably good quantitative agreement with the calculations described in Ref.~\cite{paper2} which used sophisticated numerical condensed matter simulations.  The late time features found in those calculations can be reproduced using just the gravity analogy and quantum field
theory in curved space techniques.

\acknowledgments

We would like to thank Iacopo Carusotto for helpful conversations and for sharing his numerical data with us along with an algorithm for plotting it.  P.R.A. would also like to thank Jason Bates, Greg Cook, Sarah Fisher, Michael Good, and William Kerr for helpful conversations.  This work was supported in part by the National Science Foundation under Grant Nos. PHY-0556292 and PHY-0856050.  The numerical computations herein were performed on the WFU DEAC cluster; we thank WFU��s Provost�s Office and Information Systems Department for
their generous support. A.F. thanks LPT Orsay for hospitality during various visits.

\end{document}